\documentclass[a4paper,11pt]{article}

\usepackage[T1]{fontenc}
\usepackage{lmodern}

\usepackage[margin=1in]{geometry}
\usepackage{amsmath,amssymb,amsthm}
\usepackage{enumitem}
\usepackage{tikz}

\usepackage[authoryear,round]{natbib}
\usepackage[hidelinks,breaklinks]{hyperref}

\numberwithin{equation}{section}

\setlist{
  itemsep=3pt,
  parsep=0pt,
  topsep=2pt,
  partopsep=0pt
}

\DeclareMathOperator{\CE}{CE}

\theoremstyle{plain}
\newtheorem{theorem}{Theorem}[section]
\newtheorem{proposition}[theorem]{Proposition}
\newtheorem{corollary}[theorem]{Corollary}

\theoremstyle{definition}
\newtheorem{definition}[theorem]{Definition}

\theoremstyle{remark}
\newtheorem*{remark}{Remark}

\title{Proper and Optimal Scoring Rules for Risk-Averse Agents}
\author{
	Ying Liu\thanks{Email: \texttt{y.liu12123@gmail.com}.}
	\qquad Raimundo Saona\\[0.4em]
	{\small Department of Mathematics}\\
	{\small London School of Economics and Political Science}
}
\date{November 23, 2026}

\begin{document}

\maketitle

\begin{abstract}
Scoring rules are used to elicit probabilistic beliefs through payments that depend on reports and realised outcomes. We ask how these rules should be designed to induce truthful reporting and incentivise costly information acquisition when the agent is risk averse. For agents with entropic risk preferences in finite state spaces, we characterise proper scoring rules through an exponential transformation, derive optimal structures for information acquisition under bounded payments, and obtain explicit optimal coefficients in the binary case. The risk-neutral limit preserves properness and, in the binary case, optimality, while recovering the optimal betting structure for nonbinary states. For optimal design in the binary model, it suffices to focus on rules whose associated concave function is affine on each side of the prior. The optimal coefficients depend only on the prior and the risk-aversion parameter. The binary optimal rules we construct also admit an exact transformation from a particular risk-neutral optimal rule. In the nonbinary model, optimal design reduces to choosing the best prior bound in a transformed betting mechanism.
\end{abstract}


\section{Introduction}
A decision maker often needs information that is held by another person. Eliciting this information is a classic problem in mechanism design: she must set a reward rule in advance that incentivises the information holder to share it. For example, an investor wants to know whether a project will succeed, so that she can decide whether to invest, while an analyst can conduct costly research to form a belief about the project's success. The investor therefore commits to a payment rule: the analyst will receive a monetary reward depending on both the report and the outcome of the project. The analyst then conducts research and reports the belief to the investor as a probabilistic forecast. Once the outcome is observed by both parties, the investor pays the analyst according to the rule. A payment rule of this form is called a \emph{scoring rule} \citep{GneitingRaftery2007}.

The analyst cares about their own payment rather than about helping the investor. They may therefore misreport their forecast if doing so increases their reward. A scoring rule is called \emph{proper} if it makes truthful reporting optimal for the analyst: whatever their belief is, reporting it maximises their evaluation of the resulting random payment. Under risk neutrality, this evaluation is the expected payment.

The characterisation of proper scoring rules has been studied extensively in the classic literature, under the assumption of risk neutrality. \citet{Savage1971} shows that, when the outcome space is finite, proper scoring rules correspond to convex functions of the agent's belief: a rule is proper exactly when its payments are given by the supporting affine functions of such a function. \citet{GneitingRaftery2007} further develop this representation and extend the theory of proper scoring rules to general probability spaces.

In practice, agents are often risk averse: they care not only about the expected payment but also about how risky it is. A scoring rule that is proper under risk neutrality is not necessarily proper for such agents.

A common example of such a rule is the quadratic scoring rule. Consider again an investor who wants to elicit an analyst's belief about whether a project will succeed. The project has two possible outcomes: success (\(x=1\)) and failure (\(x=0\)). The analyst reports a probability of success \(r\in[0,1]\), and the investor pays the analyst according to
\[
s(r,x)=100\bigl[1-(r-x)^2\bigr].
\]
Suppose that the analyst believes the project will succeed with probability \(0.75\). Reporting this belief truthfully yields 93.75 if the project succeeds and 43.75 if it fails, giving an expected payment of 81.25. By contrast, reporting \(0.5\) yields 75 regardless of the outcome. A risk-neutral analyst prefers the truthful report, because it gives the higher expected payment. A risk-averse analyst, however, also dislikes the risk in the truthful report, whose payment may be as low as 43.75. If the analyst is sufficiently risk averse, they value this risky random payment below the certain 75 and prefer to report \(0.5\). Hence, truthful reporting is not optimal for such an analyst, and the quadratic rule is not proper for them.

A substantial literature on belief elicitation under risk preferences addresses this issue. It can be organised around three approaches. First, some studies retain scoring rules that are proper under risk neutrality and ask how agents with different risk preferences distort their reports. \citet{PeysakhovichPlagborgMoller2012} show that risk-averse expected-utility maximisers tend to shade their reports towards the uniform distribution, as in the example above. \citet{ChambersHealyLambert2019} characterise the reports induced by proper scoring rules for more general preferences, which may combine risk aversion and ambiguity aversion.

Second, when the agent's utility function or preference model is known, another strand changes the payment rule so that truthful reporting is optimal, or develops procedures for correcting distorted reports. For expected-utility agents, \citet{FangStinchcombeWhinston2010} observe that if a scoring rule \(h\) is proper for a risk-neutral agent and takes values in the range of an invertible von Neumann--Morgenstern utility function \(u\), then the monetary scoring rule \(u^{-1}\circ h\) is proper for an agent with utility \(u\). For agents with non-expected-utility preferences, such as rank-dependent utility, \citet{OffermanEtAl2009}, \citet{KothiyalSpinuWakker2011} and \citet{Carvalho2015} develop methods for correcting reported beliefs or for recovering the decision weights that these agents report in place of their beliefs.

Third, some studies seek elicitation procedures that remain valid when the designer has only limited knowledge of the agent's risk preferences. \citet{Lambert2011} characterises probability-elicitation mechanisms that are incentive compatible for broad classes of probabilistically sophisticated agents. With deterministic event-contingent payments, such mechanisms can generally elicit only whether an event is more likely than its complement. Exact probabilities can instead be elicited by randomising over fixed monetary prizes.

These studies focus on how to make an agent report truthfully. But a proper scoring rule, which makes truthful reporting optimal, may still not pay the analyst enough to do the costly research in the first place: if the analyst's gain from being informed is smaller than the cost, the analyst has no reason to acquire the information. Moreover, different proper scoring rules can offer very different gains.

In our model, we refer to the decision maker as the principal and the information holder as the agent. We study how the principal should design the scoring rule when the agent's research is costly. There are finitely many possible states of the world. In the simplest case there are two: the project either succeeds or fails. We first study this case and then extend the analysis to any finite number of states. The research produces a signal about the state. The joint distribution of the signal and the state is known to both parties, and so is the belief that each possible signal leads to. The agent cannot choose the signal, only whether to observe it. The timing is as follows.
\begin{enumerate}
    \item The principal announces a scoring rule.

    \item The agent decides whether to pay a fixed cost to observe the signal. If they do, they observe the signal privately and update their belief; otherwise, they retain their prior belief.

    \item The agent reports a probability forecast of the state. They are free to report any forecast, not necessarily their true belief.

    \item The state is revealed, and the principal pays the agent according to the scoring rule.
\end{enumerate}

The agent knows the cost, but the principal does not. The agent is risk averse and evaluates risky payments with exponential utility, where a parameter \(\theta\), known to both parties, measures the degree of risk aversion. Every payment must lie between 0 and 1 (in some unit of money), whatever the report and the realised state. Since the principal does not know the cost, she wants the agent to be willing to do the research even when the cost is high. We therefore look for the scoring rule that maximises the highest cost at which the agent is still willing to observe the signal.

This setting raises two questions.
\begin{enumerate}
    \item Which scoring rules make truthful reporting optimal for such a risk-averse agent? This question belongs to the second approach above: the preference model is known, and we change the payment rule.

    \item Among these rules, which one gives the strongest incentive to acquire information?
\end{enumerate}

Restricting attention to proper scoring rules in the second question is without loss of generality. We only consider rules under which, at every belief, the agent has an optimal report; otherwise the reporting problem has no solution. For every such rule, the revelation principle \citep{Myerson1979} gives a proper rule that pays, at each belief, what an optimal report paid in every state under the original rule. Because the payments are preserved in every state, this holds for risk-averse agents as well. The agent's best payments, with and without the signal, are also unchanged, so the new rule gives the same incentive to acquire information. Its payments are among those of the original rule, so it satisfies the same payment bounds.

Our second question, which scoring rule gives the strongest incentive to acquire information, belongs to the literature on incentivising costly information acquisition through scoring rules and contracts. The papers in this literature differ first in how the agent acquires information and how the information is used. In the closest paper to ours, \citet{LiHartlineShanWu2025}, the agent decides whether to acquire one signal whose distribution of posteriors is known to the principal, and then reports a belief.
In other papers the agent chooses how much to learn. \citet{Osband1989} considers a forecaster who can refine a forecast at a constant marginal cost per unit of precision, and \citet{Wu2026} lets the agent choose the precision of a signal about a continuous state. In \citet{Clemen2002} and \citet{NeymanNoarovWeinberg2021}, the forecaster collects samples and updates a Bayesian belief. The expert in the former chooses the sample size in advance, before seeing any data, while the forecaster in the latter buys samples of a coin one at a time and stops adaptively.
In others the agent chooses what to learn. In \citet{HartlineShanLiWu2023} there are several tasks, each with its own cost and signal, and the agent chooses on which of them to exert effort. In \citet{WhitmeyerZhang2023} and \citet{RappoportSomma2017} the agent can choose any information structure, that is, any distribution of posteriors, at a cost; in the former the agent privately observes the posterior and reports it, while in the latter the outcome of the experiment is observable and contractible.
What the principal knows also varies. \citet{LiHartlineShanWu2025} assume that she knows the distribution of posteriors, \citet{ChenYu2026} that she knows only a set of possible information structures, and \citet{ChenWuWuYang2023} that she knows neither the costs nor the information structure and learns the scoring rule through repeated interaction. What the agent does with the information varies too. \citet{PapireddygariWaggoner2022} let the agent take a hidden action after acquiring information, so that the principal observes only the final outcome, \citet{LindbeckWeibull2020} delegate the investment decision itself to the agent, and in \citet{KeppoLi2025} the agent exerts costly due-diligence effort and then recommends whether to invest.

The papers also differ in how payments are constrained and in what the principal wants, and the optimal rule depends on both. Payments are bounded in every state in \citet{LiHartlineShanWu2025}, \citet{HartlineShanLiWu2023} and \citet{Wu2026}, bounded in expectation in \citet{ChenYu2026}, and only required to be non-negative in \citet{PapireddygariWaggoner2022}. The principal maximises the agent's gain from the signal in \citet{LiHartlineShanWu2025}, its worst case over information structures in \citet{ChenYu2026}, the precision of the forecast in \citet{NeymanNoarovWeinberg2021} and of the agent's signal in \citet{Wu2026}, and minimises her own payment in \citet{PapireddygariWaggoner2022} and \citet{WhitmeyerZhang2023}. Hence, the optimal rule has different forms. When the gain from the signal is maximised and the information structure is known, it is V-shaped \citep{LiHartlineShanWu2025,ChenYu2026}, and in \citet{LiHartlineShanWu2025} it depends on the information structure only through the prior mean. When the precision is maximised, it pays the whole budget if the report is close enough to the state and nothing otherwise \citep{Wu2026}. Several of these papers restrict attention to proper scoring rules without loss of generality, using a revelation principle \citep{LiHartlineShanWu2025,ChenWuWuYang2023,PapireddygariWaggoner2022}.

Most of these papers assume that the agent is risk neutral or, as in \citet{Wu2026}, measure the cost of information in utility, so risk attitude can be absorbed into the payments. A few papers focus on the effect of risk aversion. \citet{WhitmeyerZhang2023} show that with a risk-averse agent the principal cannot attain the first best, because inducing the agent to learn exposes the agent to risk. \citet{LindbeckWeibull2020} study a principal who delegates information acquisition and the investment decision to an agent protected by limited liability, and find numerically that the principal can benefit from a risk-averse agent, because risk aversion weakens the asymmetry between bonuses and penalties that limited liability creates. In an appendix, \citet{ClarkReggiani2021} consider a risk-averse agent, for whom the optimal distortion of incentives also provides insurance. Our problem differs from these in what the agent chooses and in what the principal optimises.

We take the bounded scoring-rule design framework of \citet{LiHartlineShanWu2025} as our starting point, in which the agent is risk neutral, and study how risk aversion changes the conditions for truthful reporting and the optimal incentives for information acquisition.
We model the agent as an expected-utility maximiser with exponential utility. For a monetary payoff \(z\in\mathbb{R}\) and a risk-aversion parameter \(\theta>0\), the utility function is
\[
u_\theta(z)=-e^{-\theta z}.
\]
This function exhibits constant absolute risk aversion (CARA), with coefficient \(\theta\). The associated certainty equivalent is called the entropic certainty equivalent (CE). We choose this model for two reasons. First, under expected utility, evaluating the payment in two steps, first given the signal and then over the possible signals, gives the same value as evaluating the final payment directly. Otherwise, the value of acquiring information would depend on whether the agent evaluates the payment step by step or as a whole. Second, adding the same amount of money to every outcome does not change how the agent ranks monetary lotteries. As a result, a fixed acquisition cost reduces the CE by exactly that cost \citep{FollmerSchied2002,FollmerWeber2015}, and the highest cost at which the agent is willing to acquire information is the difference between the CE with information and the CE without it. Among strictly risk-averse expected-utility preferences, exponential utility is the only one with the second property \citep{Muller2007}, and it recovers risk neutrality as \(\theta\to0\).

The principal wants to induce information acquisition while keeping payments bounded. Following \citet{LiHartlineShanWu2025}, we therefore require \(s(r,x)\in[0,1]\) for every report \(r\) and every realised state \(x\). The upper bound caps what the principal pays in every state, not only on average. A bound on the expected payment alone would allow very large payments in unlikely states. The lower bound is a limited-liability constraint: the agent is never asked to pay \citep{PapireddygariWaggoner2022,WhitmeyerZhang2023}. Bounded payments are also used in \citet{HartlineShanLiWu2023} and \citet{Wu2026}. In the risk-neutral problem, the bound also prevents the principal from scaling the payments up to obtain arbitrarily large incentives \citep{LiHartlineShanWu2025}. From the agent's side, the bound of 1 fixes the unit of money in which the risk-aversion parameter \(\theta\) is measured.

\paragraph{Main results.}
We first work in the binary-state setting. For an agent with risk aversion \(\theta\), we characterise the proper scoring rules (Theorem~\ref{Properness}). The characterisation parallels \citet{Savage1971}: once each payment \(s\) is transformed into \(e^{-\theta s}\), a rule is proper if and only if its transformed payments are the supporting affine functions of a positive concave function, just as under risk neutrality they are the supporting lines of a convex function. We show that, as \(\theta\to0\), the limit of proper rules is a proper rule for the risk-neutral agent (Corollary~\ref{RiskNeutralProperness}). We also characterise the rules that are proper for every \(\theta\) at once (Theorem~\ref{any-Properness}). For reports that assign positive probability to both states, these rules either make payments independent of the report or pay a fixed higher amount when the state reported as more likely occurs and a fixed lower amount otherwise. Reports expressing certainty about the state allow some additional payment choices. The first two results extend to any finite number of states (Theorem~\ref{PropernessMulti} and Corollary~\ref{RiskNeutralPropernessMulti}).

Next, in the binary case, we study the optimal rule. We show that for any informative distribution of posteriors, there is an optimal rule whose concave function in the characterisation of properness is an inverted V with its peak at the prior (Theorem~\ref{Optimality}). In particular, all reports strictly below the prior give the same payments in each state, and the same holds for reports strictly above the prior. We derive its coefficients explicitly: they depend only on the prior and on \(\theta\), not on the distribution of posteriors (Proposition~\ref{Coefficient}). As \(\theta\to0\), a family of optimal rules converges to an optimal risk-neutral rule (Corollary~\ref{RiskNeutralOptimalStructure}). We also give an explicit transformation, depending on \(\theta\), that maps a particular risk-neutral optimal rule into an optimal rule for every \(\theta\) (Proposition~\ref{Transformation}).

With finitely many states, the optimal rule can be implemented as a betting mechanism (Theorem~\ref{OptimalityMulti}), as in the risk-neutral problem \citep{LiHartlineShanWu2025}. The principal offers every bet whose payments lie in \([0,1]\) and whose expected transformed payment under the prior is at least a specified bound, and the agent chooses the bet that is best under their belief. This choice is a fractional knapsack problem. As \(\theta\to0\), the bets and the acquisition incentives converge to those of the risk-neutral mechanism, for corresponding prior bounds (Corollary~\ref{RiskNeutralOptimalStructureMulti}).

The remainder of the paper is organised as follows. Section~\ref{sec:preliminaries} sets up the model and gives a binary-state example. Section~\ref{sec:binary} presents the results for binary states and Section~\ref{sec:nonbinary} those for finitely many states. Section~\ref{sec:conclusion} concludes.

\section{Preliminaries}
\label{sec:preliminaries}
\subsection{Environment and Timing}
A principal seeks information about an uncertain state \(X\) that takes values in a finite set \(\mathcal X=\{x_1,\ldots,x_m\}\). An agent can acquire information about \(X\) at a cost, by privately observing a signal \(Y\). The principal aims to design a scoring rule with bounded payments that induces the agent to report truthfully and to find information acquisition worthwhile even when its cost is high. 

Let \(\pi\) denote the joint distribution of \(X\) and \(Y\), and let \(\Delta(\mathcal X)\) denote the set of probability distributions over \(\mathcal X\). After observing \(Y=y\), the agent forms the posterior \(q(y)\in\Delta(\mathcal X)\), with \(q_i(y)=\Pr_\pi(X=x_i\mid Y=y)\). We write \(Q=q(Y)\) for the random posterior and let \(F\) denote its distribution induced by \(\pi\). The prior is \(p\in\Delta(\mathcal X)\), with \(p_i=\Pr_\pi(X=x_i)\). Bayes plausibility gives \(\mathbb E_{Q\sim F}[Q]=p\).
We place no restrictions on \(\pi\) other than requiring the prior to have full support, so \(p_i>0\) for every \(i=1,\ldots,m\). The signal \(Y\) and the posterior \(Q\) may take infinitely many values. We call the signal informative if \(\Pr(Q\neq p)>0\).
The joint distribution \(\pi\) is common knowledge, so both parties also know the mapping \(y\mapsto q(y)\) from signals to posterior beliefs.

The principal first commits to and announces a scoring rule \(s\). For each report \(r\in\Delta(\mathcal X)\) and each realised state \(x\in\mathcal X\), the rule specifies a payment \(s(r,x)\) to the agent. After seeing the rule, the agent decides whether to incur a fixed monetary cost \(c\geq0\) to observe \(Y\). The agent knows \(c\), but the principal does not. If the agent acquires information, they privately observe the realised signal \(y\) and update their belief to \(q(y)\). Otherwise, their belief remains \(p\). The agent then submits a report \(r\in\Delta(\mathcal X)\). Once the state \(X\) is observed, the principal pays \(s(r,X)\). The agent's net monetary payoff is \(s(r,X)-c\) if they acquired information and \(s(r,X)\) otherwise.

\subsection{Preferences and Properness}
The agent evaluates monetary payoffs using exponential utility \(u_\theta(z)=-e^{-\theta z}\), where \(z\) is a monetary payoff and \(\theta>0\) is the coefficient of absolute risk aversion known to both parties. Specifically,
\[
-\frac{u_\theta''(z)}{u_\theta'(z)}=\theta.
\]
This coefficient is unchanged by positive affine transformations of the utility function, but its numerical value depends on the monetary unit. If one new monetary unit equals \(a>0\) old units, writing \(t=z/a\) gives \(\widetilde u(t)=u_\theta(at)=-e^{-a\theta t}\). The coefficient in the new unit is therefore \(a\theta\), while the agent's preferences remain unchanged.

The associated certainty equivalent is the certain monetary payoff whose utility equals the expected utility of the random payoff \citep[Section~4.4]{FollmerWeber2015}. For a bounded random payoff \(Z\), it is defined by
\[
u_\theta\bigl(\CE_\theta(Z)\bigr)
=
\mathbb E[u_\theta(Z)].
\]
For exponential utility, solving this equation gives the following expression.

\begin{definition}[Entropic certainty equivalent]
For a bounded random payoff \(Z\) and \(\theta>0\), the entropic certainty equivalent is
\[
\CE_\theta(Z)
=
-\frac{1}{\theta}\log\mathbb E\!\left[e^{-\theta Z}\right].
\]
\end{definition}

Larger values of \(\theta\) correspond to greater absolute risk aversion. We define \(\CE_0(Z):=\mathbb E[Z]\), which is the limit of \(\CE_\theta(Z)\) as \(\theta\to0^+\). Thus, \(\theta=0\) represents the risk-neutral benchmark. As \(\theta\to\infty\), \(\CE_\theta(Z)\) converges to the essential infimum of \(Z\) \citep{FollmerSchied2002}.

Let \(U_\theta(q,r)\) denote the certainty equivalent of the payment \(s(r,X)\) under belief \(q\), for \(q,r\in\Delta(\mathcal X)\). For \(\theta>0\),
\[
U_\theta(q,r)
=
-\frac{1}{\theta}\log
\left(
\sum_{i=1}^{m} q_i e^{-\theta s(r,x_i)}
\right).
\]
Under risk neutrality, \(U_0(q,r)=\sum_{i=1}^{m}q_i s(r,x_i)\).

Given belief \(q\), the agent chooses a report to maximise the certainty equivalent of their net monetary payoff. If the agent has acquired information, translation invariance implies that the fixed cost \(c\) reduces the certainty equivalent by the same amount for every report. Thus, an optimal report at belief \(q\) satisfies
\[
r\in\operatorname*{arg\,max}_{\hat r\in\Delta(\mathcal X)}
U_\theta(q,\hat r).
\]

We adapt the classical notion of a proper scoring rule \citep{Savage1971,GneitingRaftery2007} to these preferences.
\begin{definition}[\(\theta\)-proper scoring rule]
For \(\theta\geq0\), a scoring rule \(s\) is \(\theta\)-proper if
\[
U_\theta(q,q)\geq U_\theta(q,r)
\qquad
\text{for all }q,r\in\Delta(\mathcal X).
\]
\end{definition}

Throughout, properness is understood in the weak sense. Truthful reporting is an optimal choice and may tie with other reports. We impose no tie-breaking rule, so an optimal report may leave the agent's belief unidentified.

\subsection{Information Acquisition Incentives}
When deciding whether to acquire information, the agent evaluates the payment they would receive before knowing either the posterior \(Q\) or the realised state \(X\). After observing \(Q\), the agent can choose a report that maximises their conditional certainty equivalent. Without information, the agent chooses a report under the prior \(p\). For a scoring rule \(s\) under which an optimal report exists at every belief, define the acquisition incentive, before deducting the acquisition cost, by
\[
\mathcal I_\theta(s)
=
\CE_\theta\!\left(\max_{r\in\Delta(\mathcal X)} U_\theta(Q,r)\right)
-
\max_{r\in\Delta(\mathcal X)} U_\theta(p,r).
\]

By the revelation principle argument in the introduction, we can restrict attention to \(\theta\)-proper scoring rules in the optimal design problem. For such rules, truthful reporting attains both maxima. We can therefore use the truthful report \(Q\) to compute the optimal conditional certainty equivalent \(U_\theta(Q,Q)\), since all optimal reports at a given posterior yield the same conditional value. Hence, for \(\theta\)-proper scoring rules
\begin{equation}
\mathcal I_\theta(s)
=
\CE_\theta\bigl(U_\theta(Q,Q)\bigr)-U_\theta(p,p).
\label{eq:acquisition-incentive}
\end{equation}

The agent pays the fixed cost \(c\) if they acquire information. By translation invariance, the net gain from acquiring information is
\[
\begin{aligned}
\CE_\theta\bigl(U_\theta(Q,Q)-c\bigr)-U_\theta(p,p)
&=\CE_\theta\bigl(U_\theta(Q,Q)\bigr)-U_\theta(p,p)-c\\
&=\mathcal I_\theta(s)-c.
\end{aligned}
\]
The agent strictly prefers to acquire information if \(c<\mathcal I_\theta(s)\), strictly prefers not to acquire it if \(c>\mathcal I_\theta(s)\), and is indifferent at equality. The principal seeks to maximise this cost threshold subject to the payment bounds specified below.

\begin{remark}
Under risk neutrality, for proper scoring rules, \citet[Eq.~(1)]{LiHartlineShanWu2025} define the acquisition incentive as the expected payment gain from reporting the posterior rather than the prior. In our notation,
\[
\mathcal I_0(s)
=
\mathbb E_{Q\sim F}\bigl[U_0(Q,Q)-U_0(Q,p)\bigr].
\]
The difference inside the expectation compares the expected payments from reporting \(Q\) and reporting \(p\), both evaluated under the posterior \(Q\).

For \(\theta>0\), the same comparison can be expressed as a difference of certainty equivalents. Since \(U_\theta(Q,p)\) is the conditional certainty equivalent of the payment \(s(p,X)\), time consistency gives
\[
\CE_\theta\bigl(U_\theta(Q,p)\bigr)=U_\theta(p,p).
\]
Hence,
\[
\mathcal I_\theta(s)
=
\CE_\theta\bigl(U_\theta(Q,Q)\bigr)
-
\CE_\theta\bigl(U_\theta(Q,p)\bigr).
\]
The two terms must be evaluated separately, because the certainty equivalent of a difference does not in general equal the difference of the certainty equivalents.
\end{remark}

\subsection{Payment Constraints and Optimal Design}
\label{sec:theta-optimal}
We call a scoring rule \(s\) feasible at \(\theta\) if it is \(\theta\)-proper and satisfies the ex-post payment bounds
\[
0\leq s(r,x)\leq 1
\qquad
\text{for every } r\in\Delta(\mathcal X) \text{ and } x\in\mathcal X.
\]
The family of models with payment bounds \([0,1]\) and \(\theta\geq0\) covers any fixed finite payment interval \([B_L,B_U]\), where \(B_L<B_U\). A model with payments \(s'(r,x)\in[B_L,B_U]\) and risk-aversion coefficient \(\theta'\) can be normalised by setting
\[
s(r,x)=\frac{s'(r,x)-B_L}{B_U-B_L},
\qquad
\theta=\theta'(B_U-B_L).
\]
Acquisition costs and incentives are divided by \(B_U-B_L\) in the new monetary unit. This normalisation leaves the agent's optimal reporting and acquisition choices unchanged.

\begin{definition}[\(\theta\)-optimal scoring rule]
Given the distribution \(F\) of posterior beliefs and \(\theta\geq0\), a scoring rule is \(\theta\)-optimal if it maximises the acquisition incentive \(\mathcal I_\theta(s)\) over all scoring rules feasible at \(\theta\).
\end{definition}

\subsection{A Binary-State Example}
\label{sec:binary-example}
Consider a principal who asks an agent to forecast whether a project will succeed. Let \(X=1\) denote success and \(X=0\) denote failure. In this binary example, beliefs and reports are represented by the probability assigned to \(X=1\). The agent can pay a cost \(c\) to obtain a private assessment with two possible results: \(y_1\) (favourable) and \(y_2\) (unfavourable), each occurring with probability \(\alpha=\frac{1}{2}\). The corresponding posterior probabilities of success are \(q_1=\frac{3}{4}\) and \(q_2=\frac{1}{4}\), respectively. By Bayes plausibility, the prior is \(p=\frac{1}{2}\).

Consider the scoring rule
\[
s(r,x)
=
\begin{cases}
1, & r\geq\frac{1}{2},\ x=1,\\
1, & r<\frac{1}{2},\ x=0,\\
0, & \text{otherwise}.
\end{cases}
\]
Both sides of the reporting threshold offer the same two possible payments, so the agent prefers the side that gives a higher probability of receiving \(1\). At belief \(q>\frac{1}{2}\), reporting on the upper side is optimal, and at \(q<\frac{1}{2}\), reporting on the lower side is optimal. At \(q=\frac{1}{2}\), both sides give the same payment distribution. Thus, the rule is weakly \(\theta\)-proper for every \(\theta\geq0\) and satisfies the payment bounds. Truthful reporting is optimal, but every report on the same side of the threshold gives the same payments, so it need not be uniquely optimal. For example, an agent whose posterior is \(\frac{3}{4}\) receives the same payments from reporting \(0.6\), \(0.75\), or \(0.9\).

Without information, either side gives a probability of \(\frac{1}{2}\) of receiving \(1\). With information, the agent can choose the side indicated by the signal, raising this probability to \(\frac{3}{4}\). For \(\theta=\log 2\), the certainty equivalents without and with information, before deducting the cost, are respectively
\[
U_\theta(p,p)
=
-\frac{1}{\log 2}\log\!\left(
\frac{1}{2}e^{-(\log 2)\cdot0}
+
\frac{1}{2}e^{-(\log 2)\cdot1}
\right)
=
\log_2\frac{4}{3},
\]
and
\[
\begin{aligned}
\CE_\theta\bigl(U_\theta(Q,Q)\bigr)
&=
-\frac{1}{\log 2}\log\Biggl[
\frac{1}{2}\left(
\frac{3}{4}e^{-(\log 2)\cdot1}
+
\frac{1}{4}e^{-(\log 2)\cdot0}
\right)\\
&\qquad\qquad
+
\frac{1}{2}\left(
\frac{1}{4}e^{-(\log 2)\cdot0}
+
\frac{3}{4}e^{-(\log 2)\cdot1}
\right)
\Biggr]\\
&=
\log_2\frac{8}{5}.
\end{aligned}
\]
The gain from acquiring information net of its cost is therefore \(\mathcal I_{\log 2}(s)-c=\log_2(6/5)-c\). The agent strictly prefers to acquire information if \(c<\log_2(6/5)\approx0.263\) and is indifferent at equality.

Under the same scoring rule and information structure, a more risk-averse agent may be willing to incur a higher cost to acquire information. For \(\theta>0\),
\[
\mathcal I_\theta(s)
=
\frac{1}{\theta}\log\!\left[
\frac{2(1+e^{-\theta})}{1+3e^{-\theta}}
\right].
\]
We have \(\mathcal I_0(s)=\frac14\), \(\mathcal I_{\log 2}(s)\approx0.263\), and \(\mathcal I_\theta(s)\to0\) as \(\theta\to\infty\). The acquisition incentive is therefore not monotone in the risk-aversion parameter \(\theta\).

\section{Scoring Rules with Binary States}
\label{sec:binary}
As in the example in Section~\ref{sec:binary-example}, let \(\mathcal X=\{0,1\}\). We identify each belief or report with the probability assigned to \(X=1\), so that \(p\in(0,1)\) and \(q,r\in[0,1]\).

\subsection{Characterisation of \texorpdfstring{\(\theta\)}{theta}-Proper Scoring Rules}
\subsubsection{Characterisation for Fixed \texorpdfstring{\(\theta\)}{theta}}
Fix \(\theta>0\) and suppress the dependence of \(A\), \(G\), and \(g\) on \(\theta\). Define the transformed payment
\begin{equation}
A(r,x):=e^{-\theta s(r,x)}.
\label{eq:A-definition}
\end{equation}
Under belief \(q\), the certainty equivalent of the payment from report \(r\) is
\[
U_\theta(q,r)
=
-\frac{1}{\theta}\log\!\left[qA(r,1)+(1-q)A(r,0)\right].
\]
Since \(z\mapsto-\frac{1}{\theta}\log z\) is strictly decreasing, maximising the certainty equivalent is equivalent to minimising the expected transformed payment. 
For each fixed report \(r\), the expected transformed payment is affine in \(q\). Properness requires this line to attain the lower envelope at \(q=r\). This suggests a representation in terms of a concave function and its supporting lines from above.

For a concave function \(G:[0,1]\to\mathbb R\), the superdifferential of \(G\) at \(r\in[0,1]\) is
\[
\partial^+G(r)
=
\left\{a\in\mathbb R \,\middle|\, G(q)\leq G(r)+a(q-r)\ \text{for all }q\in[0,1]\right\}.
\]
Each element of \(\partial^+G(r)\) is called a supergradient of \(G\) at \(r\).

A rule \(s\) is \(\theta\)-proper exactly when the scoring rule with payments \(u_\theta(s(r,x))=-A(r,x)\) is proper under risk neutrality. The following characterisation therefore applies the Savage representation \citep{Savage1971} to these utility-transformed payments.

\begin{theorem}[Characterisation of \(\theta\)-proper scoring rules]
\label{Properness}
Fix \(\theta>0\). A scoring rule \(s\) is \(\theta\)-proper if and only if there exist a positive concave function \(G:[0,1]\to\mathbb R\) and a choice of supergradient \(g(r)\in\partial^+G(r)\) for each \(r\in[0,1]\) such that
\[
s(r,x)
=
-\frac{1}{\theta}\log\!\left[G(r)+g(r)(x-r)\right]
\]
for every \(r\in[0,1]\) and \(x\in\{0,1\}\).
\end{theorem}

\begin{proof}
\noindent\emph{Only if.}
Suppose that \(s\) is \(\theta\)-proper. Define the expected transformed payment under truthful reporting by
\begin{equation}
G(q):=qA(q,1)+(1-q)A(q,0).
\label{eq:G-definition}
\end{equation}
By \(\theta\)-properness,
\[
G(q)=\min_{r\in[0,1]}\bigl\{qA(r,1)+(1-q)A(r,0)\bigr\}.
\]
Thus \(G\) is concave as the pointwise minimum of affine functions. Also, \(G(q)>0\) because \(A(q,x)>0\) in both states.

For each fixed report \(r\), the corresponding affine function passes through \((r,G(r))\) and has slope
\[
g(r):=A(r,1)-A(r,0).
\]
Hence
\[
qA(r,1)+(1-q)A(r,0)
=
G(r)+g(r)(q-r)
\geq G(q)
\qquad\text{for all }q\in[0,1].
\]

Thus \(g(r)\in\partial^+G(r)\). Evaluating the equality at \(q=0\) and \(q=1\) gives
\begin{equation}
A(r,x)=G(r)+g(r)(x-r),
\qquad x\in\{0,1\}.
\label{eq:A-representation}
\end{equation}
Combining \eqref{eq:A-definition} and \eqref{eq:A-representation} gives the stated representation of \(s\).

\medskip
\noindent\emph{If.} Suppose that \(G\) and \(g\) satisfy the conditions in the theorem. Since \(g(r)\in\partial^+G(r)\),
\[
G(r)+g(r)(x-r)\geq G(x)>0
\qquad\text{for all }r\in[0,1],\ x\in\{0,1\}.
\]
Hence the logarithm in the representation is well defined. Exponentiating gives
\[
A(r,x)=G(r)+g(r)(x-r).
\]
Therefore,
\[
qA(r,1)+(1-q)A(r,0)
=
G(r)+g(r)(q-r)
\geq G(q),
\qquad\text{for all }q,r\in[0,1].
\]
At the truthful report \(r=q\), equality holds. Truthful reporting therefore minimises the expected transformed payment and hence maximises \(U_\theta(q,r)\). Thus \(s\) is \(\theta\)-proper.
\end{proof}

\begin{remark}
For every interior point \(r\in(0,1)\),
\[
\partial^+G(r)=[G'_+(r),G'_-(r)].
\]
At a differentiable interior point, \(g(r)=G'(r)\). At an interior kink, any value in this interval may be selected as \(g(r)\).
\end{remark}

\subsubsection{Risk-Neutral Limit}
We now let \(\theta\to0^+\) and examine how properness passes to the limiting scoring rule. The preceding characterisation applies to each fixed \(\theta>0\). Holding \(G\) and \(g\) fixed in the representation would make \(|s(r,x)|\) diverge whenever \(G(r)+g(r)(x-r)\neq1\). To study a uniformly bounded family of \(\theta\)-proper scoring rules \(s_\theta\), we therefore allow the corresponding functions \(G_\theta\) and \(g_\theta\) to vary with \(\theta\).

The transformed payments satisfy, uniformly over reports and states,
\[
A_\theta(r,x)=1-\theta s_\theta(r,x)+O(\theta^2).
\]
Consequently,
\[
\begin{aligned}
G_\theta(r)
&=1-\theta\bigl[rs_\theta(r,1)+(1-r)s_\theta(r,0)\bigr]+O(\theta^2),\\
g_\theta(r)
&=-\theta\bigl[s_\theta(r,1)-s_\theta(r,0)\bigr]+O(\theta^2).
\end{aligned}
\]

The expressions multiplying \(-\theta\) have the same algebraic form as the convex function and its selected subgradient in the characterisation of binary proper scoring rules under risk neutrality. So we define
\[
H_\theta(r):=\frac{1-G_\theta(r)}{\theta},
\qquad
h_\theta(r):=-\frac{g_\theta(r)}{\theta}.
\]

\begin{corollary}[Risk-neutral limit]
\label{RiskNeutralProperness}
Let \(\{s_\theta\}_{\theta>0}\) be a uniformly bounded family of scoring rules converging uniformly to \(s_0\) as \(\theta\to0^+\). If \(s_\theta\) is \(\theta\)-proper for every sufficiently small \(\theta>0\), then \(s_0\) is proper under risk neutrality.

More precisely, let
\[
H(r):=rs_0(r,1)+(1-r)s_0(r,0),
\qquad
h(r):=s_0(r,1)-s_0(r,0).
\]
Then \(H\) is convex, \(h(r)\) is a subgradient of \(H\) at each \(r\in[0,1]\), and
\[
s_0(r,x)=H(r)+h(r)(x-r),
\qquad r\in[0,1],\ x\in\{0,1\}.
\]
\end{corollary}

\begin{proof}
For each sufficiently small \(\theta>0\), let \(A_\theta\), \(G_\theta\), and \(g_\theta\) denote the functions associated with \(s_\theta\) in Theorem~\ref{Properness}. All expansions and convergences below are uniform over the relevant reports and states.

By the definitions of \(H_\theta\) and \(h_\theta\), the expansion of \(A_\theta\), and the uniform convergence \(s_\theta\to s_0\), as \(\theta\to0^+\) we obtain
\[
\begin{aligned}
H_\theta(r)
&=
r\frac{1-A_\theta(r,1)}{\theta}
+
(1-r)\frac{1-A_\theta(r,0)}{\theta}\\
&\longrightarrow
rs_0(r,1)+(1-r)s_0(r,0),\\[1ex]
h_\theta(r)
&=
\frac{1-A_\theta(r,1)}{\theta}
-
\frac{1-A_\theta(r,0)}{\theta}\\
&\longrightarrow
s_0(r,1)-s_0(r,0).
\end{aligned}
\]
Thus \(H_\theta\to H\) and \(h_\theta\to h\) uniformly, where \(H\) and \(h\) are defined in the corollary.

Since \(G_\theta\) is concave, \(H_\theta\) is convex. Its uniform limit \(H\) is therefore also convex. Moreover, the supergradient inequality
\[
G_\theta(q)\leq G_\theta(r)+g_\theta(r)(q-r)
\]
can be rewritten as
\[
H_\theta(q)\geq H_\theta(r)+h_\theta(r)(q-r).
\]
Taking \(\theta\to0^+\) gives
\[
H(q)\geq H(r)+h(r)(q-r)
\qquad
\text{for all }q,r\in[0,1].
\]
Thus, \(h(r)\) is a subgradient of \(H\) at \(r\).

Finally, the transformed payment representation gives
\[
\frac{1-A_\theta(r,x)}{\theta}
=
H_\theta(r)+h_\theta(r)(x-r).
\]
Since
\[
\frac{1-A_\theta(r,x)}{\theta}
=
s_\theta(r,x)+O(\theta)
\longrightarrow s_0(r,x)
\]
as \(\theta\to0^+\),
passing to the limit yields
\[
s_0(r,x)=H(r)+h(r)(x-r).
\]
By the binary Savage representation \citep[Section~3.2, eq~(12)]{GneitingRaftery2007}, a scoring rule of this form is proper under risk neutrality whenever \(H\) is convex and \(h(r)\) is a subgradient of \(H\) at each \(r\in[0,1]\). Both conditions hold, so \(s_0\) is proper under risk neutrality.
\end{proof}

\subsection{Rules That Remain Proper for Every Degree of Risk Aversion}
\begin{theorem}[Properness for every degree of risk aversion]
\label{any-Properness}
A scoring rule \(s\) is \(\theta\)-proper for every \(\theta>0\) if and only if it has one of the following forms.

\begin{enumerate}
\item There exist \(W_a,W_b\in\mathbb R\) such that
\[
(s(r,0),s(r,1))=(W_a,W_b)
\qquad
\text{for all }r\in(0,1),
\]
with endpoint payments satisfying
\[
s(0,0)=W_a,\qquad s(0,1)\leq W_b,
\]
and
\[
s(1,1)=W_b,\qquad s(1,0)\leq W_a.
\]

\item There exist \(W_L,W_H\in\mathbb R\), with \(W_L<W_H\), such that
\[
(s(r,0),s(r,1))
=
\begin{cases}
(W_H,W_L), & 0<r<1/2,\\
(W_L,W_H), & 1/2<r<1.
\end{cases}
\]
At \(r=1/2\), the payment vector may be either \((W_H,W_L)\) or \((W_L,W_H)\). The endpoint payments satisfy
\[
s(0,0)=W_H,\qquad s(0,1)\leq W_L,
\]
and
\[
s(1,1)=W_H,\qquad s(1,0)\leq W_L.
\]
\end{enumerate}
\end{theorem}

\begin{proof}
\noindent\emph{Only if.}
Suppose that the fixed scoring rule \(s\) is \(\theta\)-proper for every \(\theta>0\). We first consider the limit as \(\theta\to\infty\). For any interior belief \(q\in(0,1)\), both states have positive probability, so
\[
\lim_{\theta\to\infty}U_\theta(q,r)
=
\min\{s(r,0),s(r,1)\}.
\]
Taking this limit in the properness inequality gives
\[
\min\{s(q,0),s(q,1)\}
\geq
\min\{s(r,0),s(r,1)\}
\]
for every \(q\in(0,1)\) and \(r\in[0,1]\). Comparing any two interior reports in both directions shows that there is a common value \(W_L\) such that
\[
\min\{s(r,0),s(r,1)\}=W_L
\qquad
\text{for all }r\in(0,1).
\]

Suppose two interior reports both give payment \(W_L\) in state \(1\). Their payments in state \(0\) must coincide. Otherwise, the report with the larger payment in state \(0\) would be strictly preferred to the other report at every interior belief. The same argument applies to reports that give payment \(W_L\) in state \(0\).

Thus, interior reports generate at most two distinct payment vectors. If there is only one, write it as \((W_a,W_b)\). If there are two, they must have the form
\[
(W_{H0},W_L)
\qquad\text{and}\qquad
(W_L,W_{H1}),
\qquad W_{H0},W_{H1}>W_L.
\]

For these two vectors, dividing their expected transformed payments by \(e^{-\theta W_L}\) gives
\[
(1-q)e^{-\theta(W_{H0}-W_L)}+q
\qquad\text{and}\qquad
(1-q)+qe^{-\theta(W_{H1}-W_L)}.
\]
As \(\theta\to\infty\), these expressions converge to \(q\) and \(1-q\), respectively. Hence, when \(q<1/2\), the first vector is strictly preferred for sufficiently large \(\theta\), so truthful reporting must give \((W_{H0},W_L)\) at every such \(q\). Similarly, truthful reporting must give \((W_L,W_{H1})\) whenever \(q>1/2\).

Now fix any \(\theta>0\). For any interior beliefs \(q_-<1/2<q_+\), properness gives
\[
\begin{aligned}
(1-q_-)e^{-\theta W_{H0}}+q_-e^{-\theta W_L}
&\leq
(1-q_-)e^{-\theta W_L}+q_-e^{-\theta W_{H1}},\\
(1-q_+)e^{-\theta W_L}+q_+e^{-\theta W_{H1}}
&\leq
(1-q_+)e^{-\theta W_{H0}}+q_+e^{-\theta W_L}.
\end{aligned}
\]
By continuity of these expressions in the belief, letting \(q_-\to\left(\frac12\right)^-\) and \(q_+\to\left(\frac12\right)^+\) gives inequalities in both directions between the two limiting values. Hence,
\[
\frac12e^{-\theta W_{H0}}+\frac12e^{-\theta W_L}
=
\frac12e^{-\theta W_L}+\frac12e^{-\theta W_{H1}}.
\]
It follows that \(W_{H0}=W_{H1}=:W_H\).

Since \(r=\frac12\) is an interior report, its payment vector must be one of these two vectors. At belief \(q=\frac12\), both give the same expected transformed payment, so either may be selected at \(r=\frac12\).

It remains to determine the endpoint payments. Let \((W_a,W_b)\) denote the payment vector at interior reports sufficiently close to \(0\). Properness at belief \(0\) requires
\[
s(0,0)\geq W_a.
\]
Conversely, fix \(\theta>0\). For every sufficiently small \(q>0\), comparison of the truthful report with report \(0\) gives
\[
(1-q)e^{-\theta W_a}+qe^{-\theta W_b}
\leq
(1-q)e^{-\theta s(0,0)}+qe^{-\theta s(0,1)}.
\]
Taking \(q\to0^+\) yields \(W_a\geq s(0,0)\), so \(s(0,0)=W_a\). Substituting this equality back into the inequality gives
\[
s(0,1)\leq W_b.
\]
The symmetric argument at \(1\), using the interior payment vector near \(1\), gives
\[
s(1,1)=W_b,\qquad s(1,0)\leq W_a.
\]

In the two-vector case, the interior payment vectors near \(0\) and \(1\) are \((W_H,W_L)\) and \((W_L,W_H)\), respectively. The endpoint restrictions therefore become
\[
s(0,0)=W_H,\qquad s(0,1)\leq W_L,
\]
and
\[
s(1,1)=W_H,\qquad s(1,0)\leq W_L.
\]

\medskip
\noindent\emph{If.}
Suppose that \(s\) has one of the two stated forms. In the first case, every interior report gives the same payment vector \((W_a,W_b)\), while each endpoint report gives no larger payment in either state. Thus, truthful reporting is optimal at every interior belief. At belief \(0\), every report except \(1\) gives payment \(W_a\), while report \(1\) gives at most \(W_a\). The analogous argument applies at belief \(1\). Hence, the rule in the first case is \(\theta\)-proper for every \(\theta>0\).

In the second case, fix any \(\theta>0\). The difference between the expected transformed payments of \((W_H,W_L)\) and \((W_L,W_H)\) is
\[
(1-q)e^{-\theta W_H}+qe^{-\theta W_L}
-\bigl[(1-q)e^{-\theta W_L}+qe^{-\theta W_H}\bigr]
=
(1-2q)(e^{-\theta W_H}-e^{-\theta W_L}).
\]
For interior belief, since \(W_H>W_L\), the vector \((W_H,W_L)\) is preferred when \(0<q<\frac12\), the vector \((W_L,W_H)\) is preferred when \(\frac12<q<1\), and both give the same value when \(q=\frac12\). The endpoint reports are weakly dominated by their corresponding interior payment vectors. At beliefs \(0\) and \(1\), truthful reporting attains the maximum payment \(W_H\) in the only possible state. Hence both forms are \(\theta\)-proper for every \(\theta>0\).
\end{proof}

\subsection{Optimal Scoring Rules under Payment Bounds}
We now seek a feasible \(\theta\)-proper scoring rule that maximises the incentive to acquire information. We allow any posterior distribution \(F\) with mean \(p\in(0,1)\).

Using the acquisition incentive defined in \eqref{eq:acquisition-incentive}, we obtain
\[
\begin{aligned}
\mathcal I_\theta(s)
&=\CE_\theta\bigl(U_\theta(Q,Q)\bigr)-U_\theta(p,p)\\
&=-\frac{1}{\theta}\log\mathbb E_{Q\sim F}[G(Q)]
+\frac{1}{\theta}\log G(p)\\
&=-\frac{1}{\theta}\log
\frac{\mathbb E_{Q\sim F}[G(Q)]}{G(p)}.
\end{aligned}
\]
Since \(\theta>0\), maximising \(\mathcal I_\theta(s)\) is equivalent to minimising
\begin{equation}
\frac{\mathbb E_{Q\sim F}[G(Q)]}{G(p)}.
\label{eq:optimal-objective}
\end{equation}

The payment constraints can also be expressed using \(G\) and \(g\). Since \(A(r,x)=e^{-\theta s(r,x)}\), the bounds \(0\leq s(r,x)\leq1\) are equivalent to
\begin{equation}
e^{-\theta}
\leq
G(r)+g(r)(x-r)
\leq1
\qquad
\text{for all }r\in[0,1],\ x\in\{0,1\}.
\label{eq:transformed-payment-bounds}
\end{equation}
Thus, we minimise \eqref{eq:optimal-objective} over positive concave functions \(G\) and choices of supergradient \(g(r)\in\partial^+G(r)\) satisfying \eqref{eq:transformed-payment-bounds}.

\subsubsection{Optimal Structure}
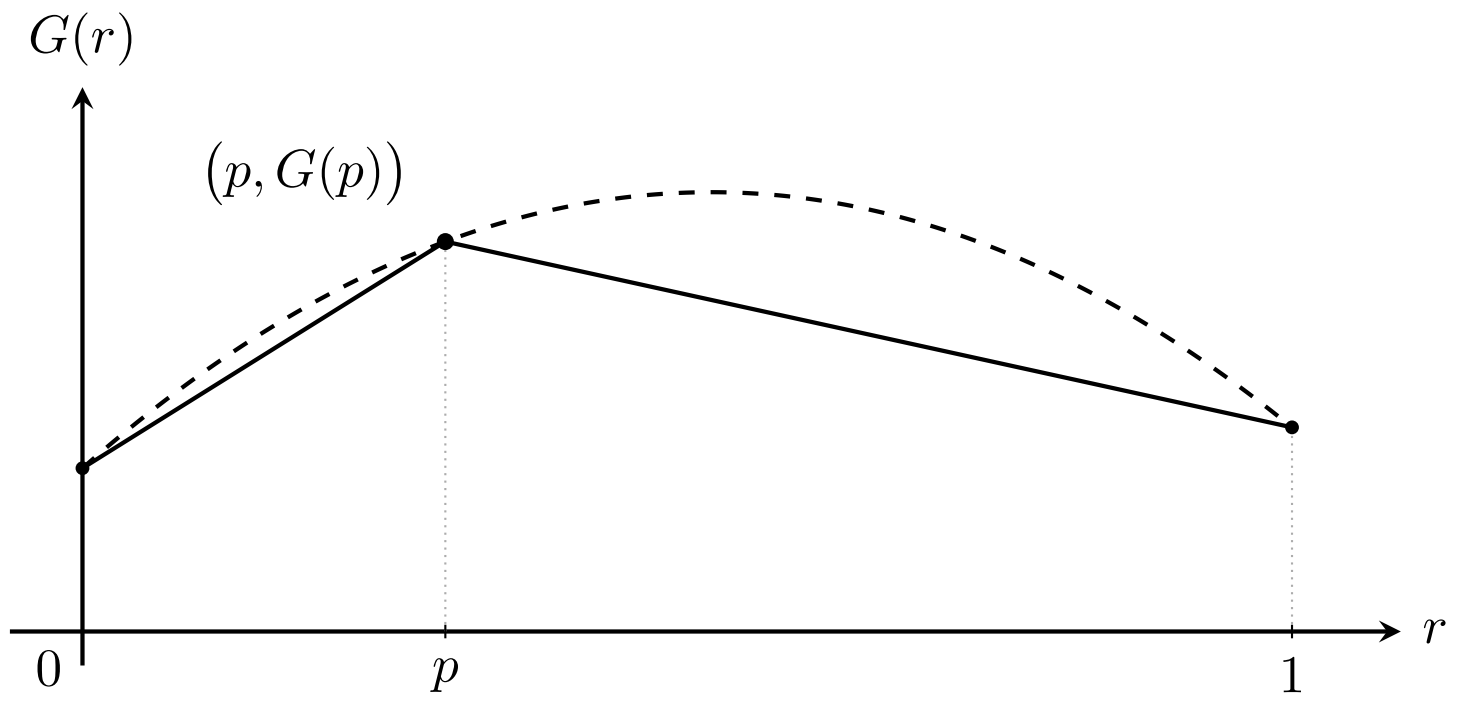
\begin{figure}[htbp]
\centering
\begin{tikzpicture}[x=8cm,y=9cm,>=stealth,font=\small]
\pgfmathsetmacro{\prior}{0.30}
\pgfmathsetmacro{\Gp}{0.12+0.78*\prior-0.75*\prior*\prior}
\coordinate (L) at (0,0.12);
\coordinate (P) at (\prior,\Gp);
\coordinate (R) at (1,0.15);
\draw[->,thick]
(-0.06,0) -- (1.09,0) node[right] {$r$};
\draw[->,thick]
(0,-0.025) -- (0,0.40) node[above] {$G(r)$};
\draw[gray!65,densely dotted] (\prior,0) -- (P);
\draw[gray!65,densely dotted] (1,0) -- (R);
\draw[thick,dashed,domain=0:1,samples=100,smooth,variable=\x]
plot ({\x},{0.12+0.78*\x-0.75*\x*\x});
\draw[thick] (L) -- (P) -- (R);
\fill (L) circle (1.3pt);
\fill (P) circle (1.6pt);
\fill (R) circle (1.3pt);
\node[above left=3pt] at (P) {$\bigl(p,G(p)\bigr)$};
\node[below left] at (0,0) {$0$};
\draw (\prior,0.005) -- (\prior,-0.005)
node[below] {$p$};
\draw (1,0.005) -- (1,-0.005)
node[below] {$1$};
\end{tikzpicture}
\caption{Construction of \(\widetilde G\) (solid) from \(G\) (dashed).}
\label{fig:inverted-v-construction}
\end{figure}

Holding \(G(p)\) fixed, replacing \(G\) by a function that lies weakly below it cannot increase the objective in \eqref{eq:optimal-objective}. Concavity places \(G\) above the chords joining its values at \(0\) and \(p\), and at \(p\) and \(1\). Replacing \(G\) by these two chords therefore weakly improves the objective. In Figure~\ref{fig:inverted-v-construction}, the dashed curve represents \(G\), while the solid segments represent \(\widetilde G\). The two functions agree at \(0\), \(p\), and \(1\), with \(\widetilde G\) lying weakly below \(G\) elsewhere.

\begin{theorem}[Optimal structure]
\label{Optimality}
Fix \(\theta>0\) and a distribution \(F\) of posterior beliefs with the prior \(p\in(0,1)\). In maximising the incentive to acquire information over feasible \(\theta\)-proper scoring rules, it is without loss of generality to restrict attention to rules whose associated function \(G\) is affine on each of \([0,p]\) and \([p,1]\).
\end{theorem}

\begin{proof}
Let \((G,g)\) be the pair associated with an arbitrary feasible \(\theta\)-proper scoring rule. Write \(v:=G(p)\) and let \(a\) and \(b\) be the slopes of the two chords:
\[
a:=\frac{G(p)-G(0)}{p},
\qquad
b:=\frac{G(1)-G(p)}{1-p}.
\]
Define
\[
\widetilde G(r)
=
\begin{cases}
v+a(r-p), & 0\leq r\leq p,\\
v+b(r-p), & p<r\leq 1.
\end{cases}
\]
The two pieces agree at \(p\), and their endpoint values are positive. Concavity of \(G\) implies \(b\leq a\), so \(\widetilde G\) is positive, continuous, and concave. It also implies \(\widetilde G(r)\leq G(r)\) for all \(r\in[0,1]\). Since \(\widetilde G(p)=G(p)\),
\[
\frac{\mathbb E_{Q\sim F}[\widetilde G(Q)]}{\widetilde G(p)}
\leq
\frac{\mathbb E_{Q\sim F}[G(Q)]}{G(p)}.
\]

By concavity of \(\widetilde G\), choose \(\gamma\in[b,a]=\partial^+\widetilde G(p)\), and set \(\widetilde g(r)=a\) for \(r<p\), \(\widetilde g(p)=\gamma\), and \(\widetilde g(r)=b\) for \(r>p\).
The associated transformed payments are
\[
\widetilde A(r,x)
=
\widetilde G(r)+\widetilde g(r)(x-r).
\]

The supergradient inequalities for \(G\) at \(0\) and \(1\) give
\[
G(p)\leq G(0)+pg(0),
\qquad
G(p)\leq G(1)+g(1)(p-1).
\]
Hence \(a\leq g(0)\) and \(g(1)\leq b\). In state \(x=0\), since \(\widetilde g(r)\in[b,a]\),
\[
\begin{aligned}
\widetilde A(r,0)&\geq v-pa=G(0)=A(0,0)\geq e^{-\theta},\\
\widetilde A(r,0)&\leq v-pb=G(1)-b\leq G(1)-g(1)=A(1,0)\leq 1.
\end{aligned}
\]
In state \(x=1\), the same slope bounds give
\[
\begin{aligned}
\widetilde A(r,1)&\geq v+(1-p)b=G(1)=A(1,1)\geq e^{-\theta},\\
\widetilde A(r,1)&\leq v+(1-p)a=G(0)+a\leq G(0)+g(0)=A(0,1)\leq 1.
\end{aligned}
\]
By Theorem~\ref{Properness}, the rule \(\widetilde s(r,x)=-\frac{1}{\theta}\log\widetilde A(r,x)\) is \(\theta\)-proper, and the bounds on \(\widetilde A\) make it feasible. The objective comparison above shows that the replacement weakly increases the incentive to acquire information. 
\end{proof}

\subsubsection{Optimal Coefficients}
By Theorem~\ref{Optimality}, it suffices to consider
\[
G(r)=
\begin{cases}
v+a(r-p), & r\leq p,\\
v+b(r-p), & r>p,
\end{cases}
\]
where \(v=G(p)\) and \(a\geq b\).

Using the supergradient selection in the proof of Theorem~\ref{Optimality}, the transformed payments take the form
\[
A(r,x)
=
\begin{cases}
v+a(x-p), & r<p,\\
v+\gamma(x-p), & r=p,\\
v+b(x-p), & r>p,
\end{cases}
\qquad
b\leq\gamma\leq a.
\]
Thus, \(A(r,1)\) is weakly decreasing in \(r\), while \(A(r,0)\) is weakly increasing. Bounding their largest and smallest values in each state gives
\begin{equation}
\begin{aligned}
v+a(1-p) &\leq1,\\
v-pb &\leq1,\\
v+b(1-p) &\geq e^{-\theta},\\
v-pa &\geq e^{-\theta},\\
b &\leq a.
\end{aligned}
\label{eq:coefficient-feasibility}
\end{equation}

\begin{proposition}[Optimal coefficients]
\label{Coefficient}
Fix \(\theta>0\) and let \(Q\sim F\) have prior \(p\in(0,1)\), with \(\Pr(Q\neq p)>0\). An optimal coefficient triple is
\[
(a^*,b^*,v^*)
=
\begin{cases}
\left(
1-e^{-\theta},
-\dfrac{p}{1-p}(1-e^{-\theta}),
e^{-\theta}+p(1-e^{-\theta})
\right),
& p\leq\dfrac12,\\[2ex]
\left(
\dfrac{1-p}{p}(1-e^{-\theta}),
-(1-e^{-\theta}),
e^{-\theta}+(1-p)(1-e^{-\theta})
\right),
& p>\dfrac12.
\end{cases}
\]
The two expressions coincide at \(p=\frac12\). Let \(G_\theta^*\) denote the function determined by these coefficients. For any choice of supergradient \(\gamma^*\in[b^*,a^*]\) at \(r=p\), the corresponding scoring rule is feasible and \(\theta\)-optimal.
\end{proposition}

\begin{proof}
By Bayes plausibility, we may define
\[
D:=
\int_{(p,1]}(q-p)\,dF(q)
=
-\int_{[0,p)}(q-p)\,dF(q)
=
\frac12\mathbb E_{Q\sim F}|Q-p|.
\]

Using the two-piece affine representation,
\[
\begin{aligned}
\mathbb E[G(Q)]
&=
v+a\int_{[0,p)}(q-p)\,dF(q)
+b\int_{(p,1]}(q-p)\,dF(q)\\
&=v-D(a-b).
\end{aligned}
\]
Since \(G(p)=v\), the objective in \eqref{eq:optimal-objective} becomes
\[
\frac{\mathbb E[G(Q)]}{G(p)}
=
1-D\frac{a-b}{v}.
\]

For an informative signal, \(\Pr(Q\neq p)>0\), \(D>0\), so minimising this ratio is equivalent to maximising
\[
\frac{a-b}{v}
\]
subject to \eqref{eq:coefficient-feasibility}. The information structure enters only through the positive factor \(D\), so the same coefficient optimisation applies to every informative distribution of posterior beliefs with prior \(p\).

The first four constraints in \eqref{eq:coefficient-feasibility} give the effective bounds
\[
a(v):=\min\left\{\frac{v-e^{-\theta}}{p},\frac{1-v}{1-p}\right\},
\qquad
b(v):=\max\left\{\frac{e^{-\theta}-v}{1-p},\frac{v-1}{p}\right\}.
\]

Since
\[
v=G(p)=pA(p,1)+(1-p)A(p,0),
\]
feasibility requires \(e^{-\theta}\leq v\leq1\). For every \(v\) in this interval,
\[
b(v)\leq0\leq a(v).
\]
Hence, the feasible coefficients for fixed \(v\) satisfy
\[
b(v)\leq b\leq a\leq a(v).
\]
The feasible region is illustrated in Figure~\ref{fig:coefficient-feasible-region}. The objective \((a-b)/v\) is maximised at its upper-left corner, where
\[
a=a(v),
\qquad
b=b(v).
\]

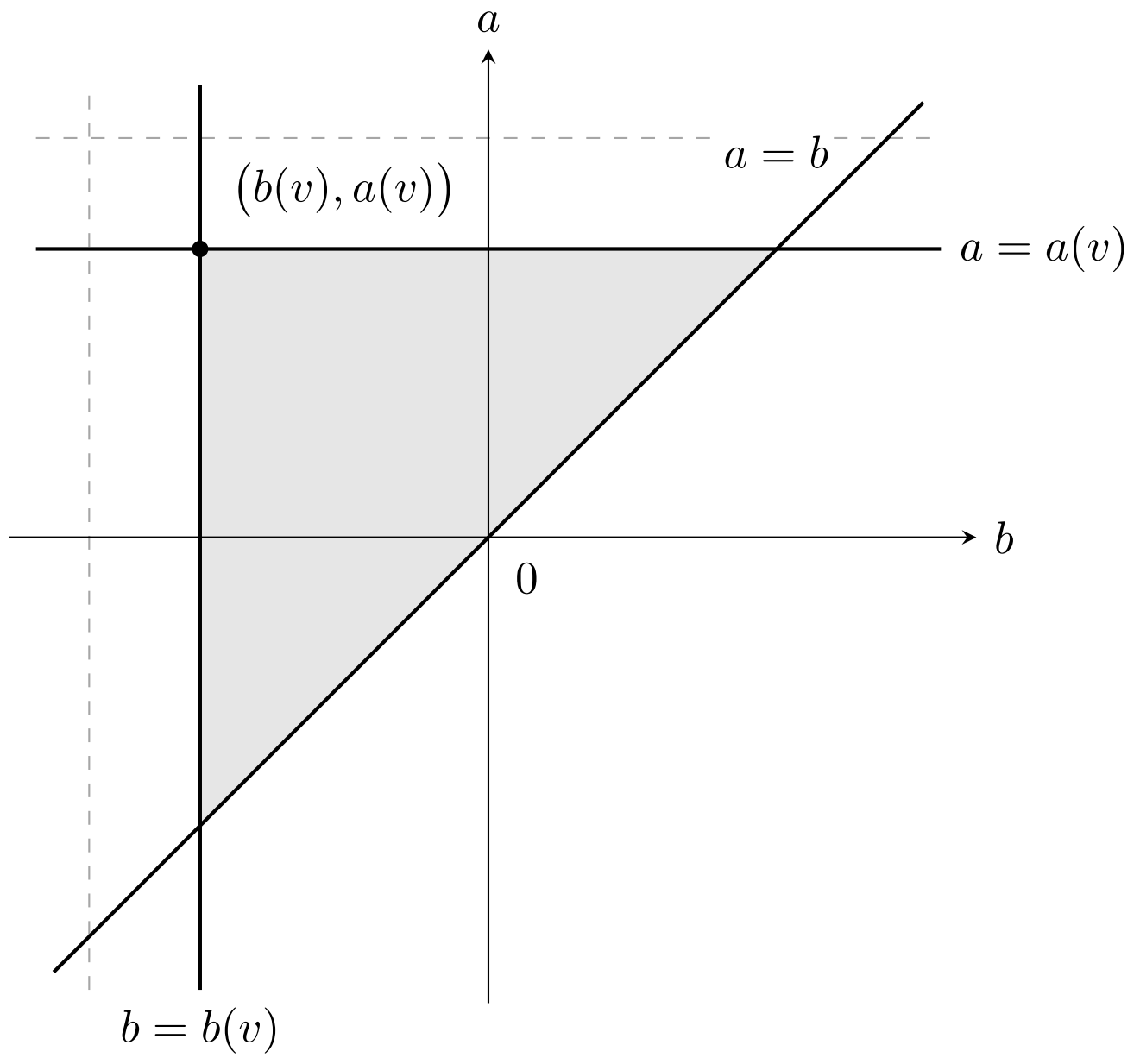
\begin{figure}[htbp]
\centering
\begin{tikzpicture}[x=3.4cm,y=3.4cm,>=stealth,font=\small]
\pgfmathsetmacro{\aUpper}{0.65}
\pgfmathsetmacro{\bLower}{-0.65}
\pgfmathsetmacro{\aRedundant}{0.90}
\pgfmathsetmacro{\bRedundant}{-0.90}
\fill[gray!20]
(\bLower,\bLower)
-- (\bLower,\aUpper)
-- (\aUpper,\aUpper)
-- cycle;
\draw[dashed,gray!70]
(-1.02,\aRedundant) -- (1.02,\aRedundant);
\draw[dashed,gray!70]
(\bRedundant,-1.02) -- (\bRedundant,1.02);
\draw[thick]
(-1.02,\aUpper) -- (1.02,\aUpper)
node[right] {$a=a(v)$};
\draw[thick]
(\bLower,-1.02) -- (\bLower,1.02);
\node[below] at (\bLower,-1.02) {$b=b(v)$};
\draw[thick] (-0.98,-0.98) -- (0.98,0.98);
\node[above left=2pt,fill=white,inner sep=1pt]
at (0.80,0.80) {$a=b$};
\draw[->]
(-1.08,0) -- (1.10,0) node[right] {$b$};
\draw[->]
(0,-1.05) -- (0,1.10) node[above] {$a$};
\node[below right=2pt] at (0,0) {$0$};
\fill (\bLower,\aUpper) circle (1.8pt);
\node[above right=3pt]
at (\bLower,\aUpper) {$\bigl(b(v),a(v)\bigr)$};
\end{tikzpicture}
\caption{Schematic feasible region in the \((b,a)\)-plane for fixed \(v\). The marked corner maximises \((a-b)/v\); dashed lines indicate redundant bounds.}
\label{fig:coefficient-feasible-region}
\end{figure}

It remains to maximise
\[
\Phi(v):=\frac{a(v)-b(v)}{v}
\qquad
\text{over }v\in[e^{-\theta},1].
\]

The two bounds defining \(a(v)\) coincide at
\[
v_a:=p+e^{-\theta}(1-p),
\]
while the two bounds defining \(b(v)\) coincide at
\[
v_b:=1-p+e^{-\theta}p.
\]
These values determine which bounds are active.

First suppose that \(p\leq\frac12\). Then \(v_a\leq v_b\), and substitution into \(\Phi\) gives
\[
\Phi(v)
=
\begin{cases}
\dfrac{v-e^{-\theta}}{v\,p(1-p)},
& e^{-\theta}\leq v\leq v_a,\\[2ex]
\dfrac{1-e^{-\theta}}{v(1-p)},
& v_a\leq v\leq v_b,\\[2ex]
\dfrac{1-v}{v\,p(1-p)},
& v_b\leq v\leq1.
\end{cases}
\]
The first expression is strictly increasing in \(v\), while the second and third are strictly decreasing. Thus, the maximum is attained at
\[
v^*=v_a=e^{-\theta}+p(1-e^{-\theta}).
\]
Substituting into \(a(v)\) and \(b(v)\) yields
\[
a^*=1-e^{-\theta},
\qquad
b^*=-\frac{p}{1-p}(1-e^{-\theta}).
\]

Now suppose that \(p>\frac12\). Then \(v_b<v_a\), and
\[
\Phi(v)
=
\begin{cases}
\dfrac{v-e^{-\theta}}{v\,p(1-p)},
& e^{-\theta}\leq v\leq v_b,\\[2ex]
\dfrac{1-e^{-\theta}}{vp},
& v_b\leq v\leq v_a,\\[2ex]
\dfrac{1-v}{v\,p(1-p)},
& v_a\leq v\leq1.
\end{cases}
\]
Again, the first expression is strictly increasing, while the remaining expressions are strictly decreasing. Therefore,
\[
v^*=v_b=e^{-\theta}+(1-p)(1-e^{-\theta}),
\]
which gives
\[
a^*=\frac{1-p}{p}(1-e^{-\theta}),
\qquad
b^*=-(1-e^{-\theta}).
\]

These coefficients satisfy \eqref{eq:coefficient-feasibility} and maximise \((a-b)/v\). Any choice of \(\gamma^*\in[b^*,a^*]\) preserves feasibility and leaves the acquisition incentive unchanged, so the corresponding scoring rule is \(\theta\)-optimal.
\end{proof}

\begin{remark}
If \(\Pr(Q=p)=1\), the acquisition incentive \(\mathcal I_\theta(s)\) is zero under every feasible \(\theta\)-proper scoring rule, so every such rule is optimal.

The constructed rule is weakly proper: truthful reporting is optimal but not uniquely optimal. All reports strictly below \(p\) give the same state-contingent payments, and the same holds for all reports strictly above \(p\).
\end{remark}

\subsubsection{Risk-Neutral Limit}
\begin{corollary}[Risk-neutral limit]
\label{RiskNeutralOptimalStructure}
For each \(\theta>0\), let \(s_\theta^*\) be the optimal rule obtained from Proposition~\ref{Coefficient} by choosing \(\gamma_\theta^*=b_\theta^*\) at \(r=p\). As \(\theta\to0^+\), \(s_\theta^*\) converges uniformly to the scoring rule
\begin{equation}
s_0^*(r,x)
=
\begin{cases}
1-x,
& p\leq\frac12,\ r<p,\\
1-p+\dfrac{p}{1-p}(x-p),
& p\leq\frac12,\ r\geq p,\\
p-\dfrac{1-p}{p}(x-p),
& p>\frac12,\ r<p,\\
x,
& p>\frac12,\ r\geq p,
\end{cases}
\label{eq:risk-neutral-optimal-limit}
\end{equation}
which is feasible and optimal under risk neutrality.
\end{corollary}

\begin{proof}
All expansions and convergences below are uniform over reports and states. With this choice at \(r=p\), we have \(A_\theta^*(r,x)=v_\theta^*+a_\theta^*(x-p)\) for \(r<p\) and \(A_\theta^*(r,x)=v_\theta^*+b_\theta^*(x-p)\) for \(r\geq p\). Substituting the optimal coefficients and factoring out \(1-e^{-\theta}\) gives
\[
\frac{1-A_\theta^*(r,x)}{\theta}
=
\frac{1-e^{-\theta}}{\theta}\,s_0^*(r,x),
\]
where \(s_0^*\) is given by \eqref{eq:risk-neutral-optimal-limit}.

Since \((1-e^{-\theta})/\theta\to1\), we have
\[
\frac{1-A_\theta^*(r,x)}{\theta}
\longrightarrow s_0^*(r,x)
\qquad\text{as }\theta\to0^+.
\]

By Proposition~\ref{Coefficient}, each \(s_\theta^*\) is \(\theta\)-proper. The uniform Taylor estimate used in the proof of Corollary~\ref{RiskNeutralProperness} yields \(s_\theta^*\to s_0^*\) uniformly, so the corollary implies that \(s_0^*\) is proper under risk neutrality. Moreover, since \(0\leq s_\theta^*(r,x)\leq1\), taking limits gives
\[
0\leq s_0^*(r,x)=\lim_{\theta\to0^+}s_\theta^*(r,x)\leq1.
\]
Hence \(s_0^*\) is feasible.

For comparison, write the risk-neutral optimal rule of
\citet{LiHartlineShanWu2025}, with the right-hand payment selected at \(r=p\), as
\[
s_{\mathrm{RN}}^*(r,x)
=
\begin{cases}
\dfrac{p}{1-p}(1-x),
& p\leq\frac12,\ r<p,\\[1ex]
x,
& p\leq\frac12,\ r\geq p,\\[1ex]
1-x,
& p>\frac12,\ r<p,\\[1ex]
\dfrac{1-p}{p}x,
& p>\frac12,\ r\geq p.
\end{cases}
\]
Comparing this rule with \eqref{eq:risk-neutral-optimal-limit} gives
\(s_0^*(r,x)=s_{\mathrm{RN}}^*(r,x)+K(x)\), where
\[
K(x)
=
\begin{cases}
\dfrac{1-2p}{1-p}(1-x), & p\leq\frac12,\\[1ex]
\dfrac{2p-1}{p}x, & p>\frac12.
\end{cases}
\]

At belief \(q\), this adds \(qK(1)+(1-q)K(0)\) to the expected payment of every report. By Bayes plausibility, the expected addition after acquiring information equals the addition at the prior:
\[
\mathbb E_{Q\sim F}\!\left[QK(1)+(1-Q)K(0)\right]
=
pK(1)+(1-p)K(0).
\]
The two rules therefore give the same acquisition incentive \(\mathcal I_0\), so \(s_0^*\) is also optimal under risk neutrality.
\end{proof}

\subsection{An Exact Transformation of Optimal Scoring Rules}
The same construction also gives an exact relation between \(s_0^*\) and \(s_\theta^*\) for every \(\theta>0\).
\begin{proposition}[Transformation of optimal scoring rules]
\label{Transformation}
For the particular risk-neutral optimal rule \(s_0^*\) constructed in \eqref{eq:risk-neutral-optimal-limit} and for every \(\theta>0\), the scoring rule
\[
s_\theta^*(r,x)
=
-\frac{1}{\theta}
\log\!\left[1-s_0^*(r,x)+s_0^*(r,x)e^{-\theta}\right]
\]
is the feasible \(\theta\)-optimal rule obtained from Proposition~\ref{Coefficient} by choosing \(\gamma_\theta^*=b_\theta^*\) at \(r=p\).
\end{proposition}

\begin{proof}
For the choice \(\gamma_\theta^*=b_\theta^*\) at \(r=p\), the identity in the proof of Corollary~\ref{RiskNeutralOptimalStructure} gives
\[
1-A_\theta^*(r,x)=(1-e^{-\theta})s_0^*(r,x).
\]
Rearranging and applying the inverse exponential transformation yields
\[
s_\theta^*(r,x)
=
-\frac{1}{\theta}\log A_\theta^*(r,x)
=
-\frac{1}{\theta}\log\!\left[1-s_0^*(r,x)+s_0^*(r,x)e^{-\theta}\right].
\]
\end{proof}

\begin{remark}
Risk-neutral optimality alone is insufficient for the transformation to yield a \(\theta\)-optimal rule. To see why the choice matters, consider substituting \(s_{\mathrm{RN}}^*\) for \(s_0^*\) in the formula. The resulting scoring rule is
\[
s_{\theta,\mathrm{RN}}(r,x)
=
-\frac{1}{\theta}\log\!\left[1-(1-e^{-\theta})s_{\mathrm{RN}}^*(r,x)\right].
\]
Since \(e^{-\theta s_{\theta,\mathrm{RN}}(r,x)}=1-(1-e^{-\theta})s_{\mathrm{RN}}^*(r,x)\), minimising the expected exponential payment is equivalent to maximising the expected payment under \(s_{\mathrm{RN}}^*\). The latter rule is proper under risk neutrality, so \(s_{\theta,\mathrm{RN}}\) is \(\theta\)-proper.
Moreover, \(0\leq s_{\mathrm{RN}}^*(r,x)\leq1\) implies
\[
e^{-\theta}\leq1-(1-e^{-\theta})s_{\mathrm{RN}}^*(r,x)\leq1.
\]
Hence \(0\leq s_{\theta,\mathrm{RN}}(r,x)\leq1\), so \(s_{\theta,\mathrm{RN}}\) is feasible.

Since \(s_0^*(r,x)=s_{\mathrm{RN}}^*(r,x)+K(x)\), the two rules satisfy
\[
e^{-\theta s_{\theta,\mathrm{RN}}(r,x)}
-
e^{-\theta s_\theta^*(r,x)}
=
(1-e^{-\theta})K(x).
\]

Let \(G_{\theta,\mathrm{RN}}\) and \(G_\theta^*\) be the functions associated with \(s_{\theta,\mathrm{RN}}\) and \(s_\theta^*\), respectively. Using the Bayes-plausibility identity established in the proof of Corollary~\ref{RiskNeutralOptimalStructure}, we obtain
\[
\begin{aligned}
\mathbb E_{Q\sim F}[G_{\theta,\mathrm{RN}}(Q)]
-\mathbb E_{Q\sim F}[G_\theta^*(Q)]
&=G_{\theta,\mathrm{RN}}(p)-G_\theta^*(p)\\
&=(1-e^{-\theta})\bigl[pK(1)+(1-p)K(0)\bigr].
\end{aligned}
\]
Thus, replacing \(s_{\theta,\mathrm{RN}}\) with \(s_\theta^*\) reduces the numerator and denominator of the objective ratio by the same amount.

From the definition of \(K\),
\[
pK(1)+(1-p)K(0)
=
\begin{cases}
1-2p, & p\leq\frac12,\\
2p-1, & p>\frac12,
\end{cases}
=
|1-2p|.
\]
When \(p\neq\frac12\) and \(\mathcal I_0(s_{\mathrm{RN}}^*)>0\), the common reduction is positive. For the transformed rule \(s_{\theta,\mathrm{RN}}\), the transformation gives
\[
\begin{aligned}
G_{\theta,\mathrm{RN}}(p)
-\mathbb E_{Q\sim F}[G_{\theta,\mathrm{RN}}(Q)]
&=1-(1-e^{-\theta})U_0(p,p)\\
&\quad-\mathbb E_{Q\sim F}\!\left[1-(1-e^{-\theta})U_0(Q,Q)\right]\\
&=(1-e^{-\theta})\left(\mathbb E_{Q\sim F}[U_0(Q,Q)]-U_0(p,p)\right)\\
&=(1-e^{-\theta})\mathcal I_0(s_{\mathrm{RN}}^*)>0,
\end{aligned}
\]
where \(U_0\) is evaluated under the original risk-neutral rule \(s_{\mathrm{RN}}^*\). Subtracting the same positive amount from both therefore strictly decreases the objective ratio, so \(s_\theta^*\) provides a strictly stronger incentive to acquire information than \(s_{\theta,\mathrm{RN}}\).
\end{remark}

The transformation in Proposition~\ref{Transformation} has a lottery interpretation. For a given report \(r\) and state \(x\), consider a lottery that pays \(1\) with probability \(s_0^*(r,x)\) and \(0\) otherwise. Its expected payment is \(s_0^*(r,x)\), and its entropic certainty equivalent is \(s_\theta^*(r,x)\).

\begin{remark}
The same transformation holds for any fixed \(\delta\in[0,1]\) if we choose \(\gamma_{\theta,\delta}^*=\delta a_\theta^*+(1-\delta)b_\theta^*\) at \(r=p\). Let \(s_{\theta,\delta}^*\) denote the resulting optimal rule. This choice affects only the payments at \(r=p\), so it suffices to verify the transformation at that report.

Its transformed payment at \(r=p\) is
\[
A_{\theta,\delta}^*(p,x)
=
v_\theta^*+\gamma_{\theta,\delta}^*(x-p),
\]
which is affine in \(\gamma_{\theta,\delta}^*\).

For \(j\in\{0,1,\delta\}\), define the limiting payment by
\[
s_{0,j}^*(p,x)
:=
\lim_{\theta\to0^+}s_{\theta,j}^*(p,x)
=
\lim_{\theta\to0^+}\frac{1-A_{\theta,j}^*(p,x)}{\theta}.
\]
These limits exist by the coefficient formulas and the uniform Taylor estimate used above. It follows that
\[
\frac{1-A_{\theta,\delta}^*(p,x)}{\theta}
=
\delta\frac{1-A_{\theta,1}^*(p,x)}{\theta}
+(1-\delta)\frac{1-A_{\theta,0}^*(p,x)}{\theta}.
\]
Taking \(\theta\to0^+\) gives
\[
s_{0,\delta}^*(p,x)
=
\delta s_{0,1}^*(p,x)+(1-\delta)s_{0,0}^*(p,x).
\]

The choices \(j=0\) and \(j=1\) give the right- and left-hand payment vectors, respectively. The identities established for these two vectors therefore give
\[
\begin{aligned}
A_{\theta,\delta}^*(p,x)
&=\delta A_{\theta,1}^*(p,x)+(1-\delta)A_{\theta,0}^*(p,x)\\
&=\delta\bigl[1-(1-e^{-\theta})s_{0,1}^*(p,x)\bigr]\\
&\quad +(1-\delta)\bigl[1-(1-e^{-\theta})s_{0,0}^*(p,x)\bigr]\\
&=1-(1-e^{-\theta})s_{0,\delta}^*(p,x).
\end{aligned}
\]

Then
\[
\begin{aligned}
s_{\theta,\delta}^*(p,x)
&=-\frac{1}{\theta}\log A_{\theta,\delta}^*(p,x)\\
&=-\frac{1}{\theta}\log\!\left[1-(1-e^{-\theta})s_{0,\delta}^*(p,x)\right].
\end{aligned}
\]

For \(r\neq p\), define \(s_{0,\delta}^*(r,x)=s_0^*(r,x)\). The corresponding payments \(s_{\theta,\delta}^*(r,x)\) are unchanged, so the transformation holds for every report and state. The same limit argument as in Corollary~\ref{RiskNeutralOptimalStructure} yields \(s_{\theta,\delta}^*\to s_{0,\delta}^*\) uniformly as \(\theta\to0^+\).
\end{remark}

\section{Scoring Rules with Nonbinary States}
\label{sec:nonbinary}
We return to the finite-state setting of Section~\ref{sec:preliminaries}, with \(\mathcal X=\{x_1,\ldots,x_m\}\) and \(p,q,r\in\Delta(\mathcal X)\).

\subsection{Characterisation of \texorpdfstring{\(\theta\)}{theta}-Proper Scoring Rules}
\subsubsection{Characterisation for Fixed \texorpdfstring{\(\theta\)}{theta}}
The binary characterisation extends to nonbinary states, with supporting lines replaced by supporting hyperplanes. Let \(e_i\) denote the \(i\)-th unit vector in \(\mathbb R^m\). For a concave function \(G:\Delta(\mathcal X)\to\mathbb R\), its superdifferential at \(r\in\Delta(\mathcal X)\) is defined by
\[
\partial^+G(r)
=
\left\{a\in\mathbb R^m\,\middle|\,
G(q)\leq G(r)+a\cdot(q-r)
\ \text{for all }q\in\Delta(\mathcal X)\right\}.
\]

\begin{theorem}[Characterisation of \(\theta\)-proper scoring rules with nonbinary states]
\label{PropernessMulti}
Fix \(\theta>0\). A scoring rule \(s\) is \(\theta\)-proper if and only if there exist a positive concave function \(G:\Delta(\mathcal X)\to\mathbb R\) and a choice of supergradients \(g(r)\in\partial^+G(r)\) such that
\[
s(r,x_i)=-\frac{1}{\theta}\log\!\left[G(r)+g(r)\cdot(e_i-r)\right],
\qquad r\in\Delta(\mathcal X),\quad i=1,\ldots,m.
\]
\end{theorem}

\begin{proof}
Write
\[
A(r,x_i):=e^{-\theta s(r,x_i)},
\qquad
A(r):=(A(r,x_1),\ldots,A(r,x_m)).
\]
As in the binary case, maximising \(U_\theta(q,r)\) over \(r\) is equivalent to minimising the expected transformed payment \(q\cdot A(r)\).

\noindent\emph{Only if.}
Suppose that \(s\) is \(\theta\)-proper and define
\[
G(q):=q\cdot A(q).
\]
Then \(G(q)>0\), and properness gives
\[
G(q)=\min_{r\in\Delta(\mathcal X)}q\cdot A(r).
\]
Since each function \(q\mapsto q\cdot A(r)\) is affine in \(q\), their pointwise minimum \(G\) is concave.

For each report \(r\), the affine function \(q\mapsto q\cdot A(r)\) supports \(G\) from above at \(r\). Let \(g(r)\in\partial^+G(r)\) denote a slope vector of this supporting affine function on the simplex, so that
\[
q\cdot A(r)
=
G(r)+g(r)\cdot(q-r)
\geq G(q)
\qquad
\text{for all }q\in\Delta(\mathcal X).
\]

Combining this affine representation with the definition of \(G(r)\), we have
\[
\begin{aligned}
G(r)+g(r)\cdot(q-r)&=\sum_{i=1}^m q_iA(r,x_i),\\
G(r)&=\sum_{i=1}^m r_iA(r,x_i).
\end{aligned}
\]
Subtracting the second equation from the first gives
\[
(A(r)-g(r))\cdot(q-r)=0
\qquad
\text{for all }q\in\Delta(\mathcal X).
\]
Hence \(A(r)=g(r)+\lambda(r)\mathbf 1\), where \(\lambda(r)\) is a scalar independent of the state index \(i\) and \(\mathbf 1=(1,\ldots,1)\). Using \(r\cdot\mathbf 1=1\), we obtain
\[
\begin{aligned}
G(r)&=r\cdot A(r)=r\cdot g(r)+\lambda(r),\\
\lambda(r)&=G(r)-r\cdot g(r).
\end{aligned}
\]
Substituting back gives
\[
\begin{aligned}
A(r)&=g(r)+\bigl(G(r)-r\cdot g(r)\bigr)\mathbf 1,\\
A(r,x_i)&=g_i(r)+G(r)-r\cdot g(r)\\
&=G(r)+g(r)\cdot(e_i-r).
\end{aligned}
\]
Since \(A(r,x_i)=e^{-\theta s(r,x_i)}\), taking logarithms yields the stated representation of \(s\).

\medskip
\noindent\emph{If.}
Conversely, suppose that \(G\) and \(g\) satisfy the conditions in the theorem. The supergradient inequality gives
\[
G(r)+g(r)\cdot(e_i-r)\geq G(e_i)>0,
\]
so the logarithm is well defined. Exponentiating the representation and using \(\sum_i q_i=1\) and \(\sum_i q_i e_i=q\), we obtain
\[
\begin{aligned}
q\cdot A(r)
&=\sum_{i=1}^m q_i\bigl[G(r)+g(r)\cdot(e_i-r)\bigr]\\
&=G(r)+g(r)\cdot(q-r)\\
&\geq G(q)=q\cdot A(q).
\end{aligned}
\]
Thus, truthful reporting minimises the expected transformed payment and hence maximises \(U_\theta(q,r)\). Therefore, \(s\) is \(\theta\)-proper.
\end{proof}

The supergradient selection is not unique. For any scalar \(\kappa(r)\), replacing \(g(r)\) by \(g(r)+\kappa(r)\mathbf1\) leaves both the supergradient inequality and the scoring rule unchanged, since \(\mathbf1\cdot(q-r)=0\) for all \(q,r\in\Delta(\mathcal X)\).

\subsubsection{Risk-Neutral Limit}
The normalisation used in the binary case also applies to nonbinary states. Define
\[
H_\theta(r):=\frac{1-G_\theta(r)}{\theta},
\qquad
h_\theta(r):=-\frac{g_\theta(r)}{\theta}.
\]

\begin{corollary}[Risk-neutral limit]
\label{RiskNeutralPropernessMulti}
Let \(\{s_\theta\}_{\theta>0}\) be a uniformly bounded family of scoring rules converging uniformly to \(s_0\) as \(\theta\to0^+\). If \(s_\theta\) is \(\theta\)-proper for every sufficiently small \(\theta>0\), then \(s_0\) is proper under risk neutrality.

More precisely, for each sufficiently small \(\theta>0\), choose the representing supergradient \(g_\theta(r):=A_\theta(r)-\mathbf1\). Let
\[
H(r):=\sum_{i=1}^m r_i s_0(r,x_i),
\qquad
h(r):=\bigl(s_0(r,x_1),\ldots,s_0(r,x_m)\bigr).
\]
Then \(H\) is convex, \(h(r)\) is a subgradient of \(H\) at each \(r\in\Delta(\mathcal X)\), and
\[
s_0(r,x_i)=H(r)+h(r)\cdot(e_i-r),
\qquad r\in\Delta(\mathcal X),\quad i=1,\ldots,m.
\]
\end{corollary}

\begin{proof}
Using the uniform Taylor expansion established in the binary case and the uniform convergence \(s_\theta\to s_0\), as \(\theta\to0^+\) we obtain
\[
\begin{aligned}
H_\theta(r)
&=\sum_{i=1}^m r_i\frac{1-A_\theta(r,x_i)}{\theta}\\
&=\sum_{i=1}^m r_i s_\theta(r,x_i)+O(\theta)
\longrightarrow H(r),\\[1ex]
h_{\theta,i}(r)
&=\frac{1-A_\theta(r,x_i)}{\theta}\\
&=s_\theta(r,x_i)+O(\theta)
\longrightarrow s_0(r,x_i)=h_i(r),
\qquad i=1,\ldots,m.
\end{aligned}
\]
Thus \(H_\theta\to H\) and \(h_\theta\to h\) uniformly.

Since \(G_\theta\) is concave, \(H_\theta\) is convex, and its limit \(H\) is also convex. Moreover, the supergradient inequality becomes
\[
H_\theta(q)\geq H_\theta(r)+h_\theta(r)\cdot(q-r).
\]
Taking \(\theta\to0^+\) gives
\[
H(q)\geq H(r)+h(r)\cdot(q-r)
\qquad
\text{for all }q,r\in\Delta(\mathcal X).
\]
Hence \(h(r)\) is a subgradient of \(H\) at \(r\). 

Finally, the transformed payment representation gives
\[
\frac{1-A_\theta(r,x_i)}{\theta}
=
H_\theta(r)+h_\theta(r)\cdot(e_i-r).
\]
Passing to the limit yields
\[
s_0(r,x_i)=H(r)+h(r)\cdot(e_i-r).
\]
By the finite-state Savage representation \citep[Section~3.1, eq~(10)]{GneitingRaftery2007}, this representation with a convex \(H\) and selected subgradients \(h(r)\) implies that \(s_0\) is proper under risk neutrality.
\end{proof}

\subsection{Structure of the \texorpdfstring{\(\theta\)}{theta}-Optimal Scoring Rules}
\subsubsection{Transformed Betting Mechanisms}
The binary construction lowers \(G\) while preserving \(G(p)\) and the payment bounds. Extending this construction to nonbinary states requires controlling the values of each supporting hyperplane at every vertex of the simplex. We therefore work directly with transformed payment vectors satisfying the bounds.

For any feasible rule, properness at the prior implies
\[
p\cdot A(r)\geq p\cdot A(p)=G(p)
\qquad
\text{for every }r\in\Delta(\mathcal X).
\]
Writing \(v:=G(p)\), every report therefore offers a vector in
\[
\left\{
A\in[e^{-\theta},1]^m
\,\middle|\,
p\cdot A\geq v
\right\}.
\]
At belief \(q\), the agent chooses a report that minimises \(q\cdot A(r)\), thereby selecting the corresponding vector \(A(r)\). We therefore consider a mechanism that lets the agent choose a transformed payment vector directly from the entire set above. This enlarges the available menu at posterior beliefs while keeping the minimum expected transformed payment at the prior equal to \(v\). We call this a transformed betting mechanism.

For each \(v\in[e^{-\theta},1]\) and each belief \(q\in\Delta(\mathcal X)\), select a vector
\[
A_v(q)\in
\operatorname*{arg\,min}_{\substack{
A\in[e^{-\theta},1]^m\\
p\cdot A\geq v
}}
q\cdot A.
\]
These choices define a vector-valued function \(A_v:\Delta(\mathcal X)\to[e^{-\theta},1]^m\). Let \(A_{v,i}(r)\) denote the \(i\)-th component of \(A_v(r)\). The induced scoring rule is
\[
s_v(r,x_i):=-\frac{1}{\theta}\log A_{v,i}(r),
\qquad r\in\Delta(\mathcal X),\quad i=1,\ldots,m.
\]

\begin{theorem}[Optimal structure with nonbinary states]
\label{OptimalityMulti}
Fix \(\theta>0\) and a distribution \(F\) of posterior beliefs with prior \(p\). Then, with \(s\) ranging over all feasible scoring rules,
\[
\max_s \mathcal I_\theta(s)
=
\max_{v\in[e^{-\theta},1]}\mathcal I_\theta(s_v).
\]
\end{theorem}

Theorem~\ref{OptimalityMulti} shows that, in maximising the acquisition incentive over feasible scoring rules, it is without loss of generality to restrict attention to this family of rules induced by transformed betting mechanisms.

For a fixed prior bound \(v\), the agent chooses an optimal transformed payment vector as follows. At the prior belief \(p\), the agent minimises \(p\cdot A\). Hence, any feasible vector satisfying \(p\cdot A=v\) is optimal.

At a posterior belief \(q\), the agent minimises \(q\cdot A\). This is a fractional knapsack problem: starting from \(A_i=e^{-\theta}\) for all \(i\), increasing component \(i\) raises \(q\cdot A\) at a rate of \(q_i/p_i\) per unit increase in \(p\cdot A\). An optimal choice therefore increases the components in nondecreasing order of \(q_i/p_i\) until \(p\cdot A=v\).

Relabel the states so that
\[
\frac{q_1}{p_1}
\leq
\frac{q_2}{p_2}
\leq
\cdots
\leq
\frac{q_m}{p_m}.
\]
Let \(k\) be the smallest index such that
\[
\sum_{i=1}^k p_i
+
\sum_{i=k+1}^m e^{-\theta}p_i
\geq v.
\]
An optimal bet assigns transformed payment \(1\) to states \(1,\ldots,k-1\), transformed payment \(e^{-\theta}\) to states \(k+1,\ldots,m\), and an intermediate transformed payment to state \(k\) so that the prior constraint binds. That is,
\[
A_{v,i}(q)
=
\begin{cases}
1,
& i<k,\\[1ex]
\displaystyle
\frac{v-\sum_{j=1}^{k-1}p_j-\sum_{j=k+1}^{m}e^{-\theta}p_j}{p_k},
& i=k,\\[2ex]
e^{-\theta},
& i>k.
\end{cases}
\]

\begin{proof}[Proof of Theorem~\ref{OptimalityMulti}]
Let \(s\) be any feasible scoring rule, write \(A(r)\) for its transformed payment vector, and set \(v:=p\cdot A(p)=G(p)\).

Properness at the prior and the payment bounds give
\[
p\cdot A(r)\geq p\cdot A(p)=v,
\qquad
A(r)\in[e^{-\theta},1]^m
\qquad
\text{for every }r\in\Delta(\mathcal X).
\]
Thus every vector offered by the original rule is available under the transformed betting mechanism. For every posterior \(q\),
\[
q\cdot A(q)
\geq
\min_{\substack{
A\in[e^{-\theta},1]^m\\
p\cdot A\geq v
}}
q\cdot A
=
q\cdot A_v(q).
\]
Taking expectations gives
\[
\mathbb E_{Q\sim F}[Q\cdot A(Q)]
\geq
\mathbb E_{Q\sim F}[Q\cdot A_v(Q)].
\]

Meanwhile, at belief \(p\), the agent chooses
\[
A_v(p)\in
\operatorname*{arg\,min}_{\substack{
A\in[e^{-\theta},1]^m\\
p\cdot A\geq v
}}
p\cdot A.
\]
The minimum is \(v\), since the original vector \(A(p)\) is available and attains this value. Hence,
\[
p\cdot A_v(p)=v=p\cdot A(p).
\]
Consequently,
\[
\begin{aligned}
&\log\mathbb E_{Q\sim F}[Q\cdot A(Q)]
-\log\!\bigl[p\cdot A(p)\bigr]\\
&\qquad\geq
\log\mathbb E_{Q\sim F}[Q\cdot A_v(Q)]
-\log\!\bigl[p\cdot A_v(p)\bigr].
\end{aligned}
\]
Since \(\theta>0\), multiplying both sides by \(-1/\theta\) shows that the transformed betting mechanism weakly increases the incentive to acquire information relative to \(s\).

To establish properness, define the associated function
\[
G_v(q)
:=
q\cdot A_v(q)
=
\min_{\substack{
A\in[e^{-\theta},1]^m\\
p\cdot A\geq v
}}
q\cdot A.
\]
Since each function \(q\mapsto q\cdot A\) is affine, their pointwise minimum \(G_v\) is concave. Moreover, every report \(r\) selects an available vector \(A_v(r)\), so
\[
q\cdot A_v(r)\geq G_v(q)=q\cdot A_v(q)
\qquad
\text{for all }q,r\in\Delta(\mathcal X).
\]
Truthful reporting therefore minimises the expected transformed payment and hence maximises \(U_\theta(q,r)\). Thus \(s_v\) is \(\theta\)-proper.

Finally, the transformed payment bounds imply
\[
0\leq s_v(r,x_i)
=-\frac{1}{\theta}\log A_{v,i}(r)
\leq1
\qquad
\text{for all }r\in\Delta(\mathcal X),\ i=1,\ldots,m.
\]
Hence \(s_v\) is feasible for every \(v\in[e^{-\theta},1]\). 

The greedy formula above shows that \(G_v(q)\) is continuous in \(v\) for every \(q\). Since \(e^{-\theta}\leq G_v(q)\leq1\), bounded convergence implies that \(\mathcal I_\theta(s_v)\) is continuous in \(v\). Hence, its maximum over the compact interval \([e^{-\theta},1]\) is attained. Together with the feasibility of every \(s_v\) and the weak improvement established above for every feasible \(s\), this proves the stated equality.
\end{proof}

\subsubsection{Risk-Neutral Limit}
Under risk neutrality, with nonbinary states, an optimal structure is the betting mechanism described in \citet[Definition~12 and Theorem~5]{LiHartlineShanWu2025}. It allows the agent to choose a payment vector \(s\in[0,1]^m\) satisfying
\[
p\cdot s\leq\beta.
\]
Obtaining an optimal rule within this family requires optimising \(\beta\). Here we fix an arbitrary \(\beta\in[0,1]\).

At belief \(q\), the agent maximises \(q\cdot s\). Specifically, relabel the states so that
\[
\frac{q_1}{p_1}
\geq
\frac{q_2}{p_2}
\geq
\cdots
\geq
\frac{q_m}{p_m}.
\]
Let \(k\) be the smallest index such that
\[
\sum_{i=1}^k p_i\geq\beta.
\]
An optimal choice defines the induced scoring rule
\[
s_{\beta,0}(q,x_i)
=
\begin{cases}
1,
& i<k,\\[1ex]
\displaystyle
\frac{\beta-\sum_{j=1}^{k-1}p_j}{p_k},
& i=k,\\[2ex]
0,
& i>k.
\end{cases}
\]

To compare this mechanism with the transformed betting mechanism, define the transformed prior bound corresponding to this fixed \(\beta\) by
\[
v_\theta:=1-(1-e^{-\theta})\beta.
\]

\begin{corollary}[Risk-neutral limit]
\label{RiskNeutralOptimalStructureMulti}
Let \(s_{\beta,0}\) be the scoring rule induced by the risk-neutral betting mechanism with parameter \(\beta\). For each \(\theta>0\), let \(s_\theta\) be induced by the transformed betting mechanism with parameter \(v_\theta\). Use the greedy selections in both mechanisms under the same nonincreasing ordering of \(q_i/p_i\), resolving ties by a fixed state order independent of \(\theta\).

As \(\theta\to0^+\), \(s_\theta\) converges uniformly to \(s_{\beta,0}\) over reports and states, and
\[
\mathcal I_\theta(s_\theta)\longrightarrow\mathcal I_0(s_{\beta,0}).
\]
\end{corollary}

\begin{proof}
Fix a belief \(q\). For this comparison, relabel the states in nonincreasing order of \(q_i/p_i\), reversing the sorting direction used above:
\[
\frac{q_1}{p_1}
\geq
\frac{q_2}{p_2}
\geq
\cdots
\geq
\frac{q_m}{p_m}.
\]
With this ordering, the boundary index under risk aversion is the smallest \(k\) satisfying
\[
e^{-\theta}\sum_{i=1}^k p_i+\sum_{i=k+1}^m p_i
\leq v_\theta.
\]
Substituting the definition of \(v_\theta\), this condition becomes
\[
1-(1-e^{-\theta})\sum_{i=1}^k p_i
\leq
1-(1-e^{-\theta})\beta,
\]
or equivalently,
\[
\sum_{i=1}^k p_i\geq\beta.
\]
Its smallest solution is exactly the boundary index under risk neutrality. Hence the two mechanisms use the same \(k\).

The selected payment vector under risk neutrality is
\[
s_{\beta,0}(q,x_i)
=
\begin{cases}
1,
& i<k,\\[1ex]
\displaystyle
\frac{\beta-\sum_{j=1}^{k-1}p_j}{p_k},
& i=k,\\[2ex]
0,
& i>k.
\end{cases}
\]
Under risk aversion, the selected transformed payment vector is
\[
A_\theta(q,x_i)
=
\begin{cases}
e^{-\theta},
& i<k,\\[1ex]
\displaystyle
\frac{v_\theta-\sum_{j=1}^{k-1}e^{-\theta}p_j-\sum_{j=k+1}^{m}p_j}{p_k},
& i=k,\\[2ex]
1,
& i>k.
\end{cases}
\]
Thus, for \(i<k\), the monetary payments under both mechanisms equal \(1\), while for \(i>k\), they equal \(0\).

For the boundary state, substituting \(v_\theta\) gives
\[
\begin{aligned}
A_\theta(q,x_k)
&=
\frac{1-(1-e^{-\theta})\beta
-e^{-\theta}\sum_{j=1}^{k-1}p_j
-\sum_{j=k+1}^{m}p_j}{p_k}\\
&=
1-(1-e^{-\theta})
\frac{\beta-\sum_{j=1}^{k-1}p_j}{p_k}\\
&=
1-(1-e^{-\theta})s_{\beta,0}(q,x_k).
\end{aligned}
\]
The same identity holds for all states and, since \(q\) is arbitrary, for every report \(r\).
Consequently,
\[
\begin{aligned}
s_\theta(r,x_i)
&=-\frac1\theta
\log\!\left[1-(1-e^{-\theta})s_{\beta,0}(r,x_i)\right]\\
&=\frac{1-e^{-\theta}}{\theta}s_{\beta,0}(r,x_i)+O(\theta)
\longrightarrow s_{\beta,0}(r,x_i).
\end{aligned}
\]
Since \(0\leq s_{\beta,0}(r,x_i)\leq1\), the remainder is uniform over reports and states. Hence \(s_\theta\to s_{\beta,0}\) uniformly.

Finally, write \(s_{\beta,0}(q):=(s_{\beta,0}(q,x_1),\ldots,s_{\beta,0}(q,x_m))\). The payment identity gives
\[
\begin{aligned}
\frac1\theta\log[p\cdot A_\theta(p)]
&=
\frac1\theta
\log\!\left[1-(1-e^{-\theta})p\cdot s_{\beta,0}(p)\right]\\
&=
-\frac{1-e^{-\theta}}{\theta}\,p\cdot s_{\beta,0}(p)+O(\theta)\\
&\longrightarrow -p\cdot s_{\beta,0}(p),
\end{aligned}
\]
and
\[
\begin{aligned}
-\frac1\theta
\log\mathbb E_{Q\sim F}[Q\cdot A_\theta(Q)]
&=
-\frac1\theta
\log\!\left[
1-(1-e^{-\theta})
\mathbb E_{Q\sim F}[Q\cdot s_{\beta,0}(Q)]
\right]\\
&=
\frac{1-e^{-\theta}}{\theta}
\mathbb E_{Q\sim F}[Q\cdot s_{\beta,0}(Q)]+O(\theta)\\
&\longrightarrow
\mathbb E_{Q\sim F}[Q\cdot s_{\beta,0}(Q)].
\end{aligned}
\]
Adding these limits yields
\[
\lim_{\theta\to0^+}\mathcal I_\theta(s_\theta)
=
\mathbb E_{Q\sim F}[Q\cdot s_{\beta,0}(Q)]
-p\cdot s_{\beta,0}(p)
=
\mathcal I_0(s_{\beta,0}).
\]
\end{proof}

\section{Conclusion}
\label{sec:conclusion}
This paper derives explicit optimal rules for binary states and reduces optimal design for nonbinary states to choosing a prior bound. The constructed rules are weakly proper, so optimal reports need not identify the posterior belief. Further work could study information acquisition under strictly proper scoring rules. Another direction is to determine the optimal prior bound in the nonbinary-state model.

\section*{Author Contributions}
Ying Liu (Y.L.) carried out most of the research. Y.L. developed the model, derived and proved the results, surveyed the literature, wrote the manuscript, and
independently proposed some of the directions the paper pursues.
Raimundo Saona (R.S.) conceived and supervised the project. R.S. proposed extending the scoring-rule design framework of \citet{LiHartlineShanWu2025} to risk-averse agents with entropic risk preferences, posed the research questions, and provided directions and feedback throughout, including comments on drafts of
the manuscript.

\section*{Generative AI Statement}
We used ChatGPT to suggest improvements to the language, notation, LaTeX and presentation of drafts and to identify potentially relevant literature. 
We used Claude to translate occasional phrases and suggest wording revisions in parts of the introduction and preliminaries. 
After drafting the mathematical arguments, we used ChatGPT to review them for possible omissions and unclear exposition, and checked any resulting changes against the definitions and our calculations before including them.
All references included in the paper were checked against the original sources.

\bibliographystyle{plainnat}
\bibliography{template-BibFile}

\end{document}